%% file: main.tex
\documentclass[12pt]{article}
\usepackage{hyperref} 

\hypersetup{
    colorlinks,
    linkcolor={red!50!black},
    citecolor={green!50!black},
    urlcolor={blue!80!black}
}
\usepackage{comment}
\usepackage{xcolor} 
\usepackage{amsmath, amssymb, amsthm}
\usepackage{mathrsfs} 
\usepackage{graphicx,psfrag,epsf}
\usepackage{enumerate}
\usepackage[utf8]{inputenc}
\usepackage{float} 

\usepackage{ulem}

\usepackage{enumitem}

\usepackage{longtable} 

\usepackage{newunicodechar}
\newunicodechar{ℝ}{\mathbb{R}}
\usepackage{algorithm}
\usepackage{algorithmicx}
\usepackage{array}
\usepackage{multirow} 
\usepackage{cleveref} 
\crefformat{equation}{(#2#1#3)}

\usepackage{algcompatible}

\theoremstyle{plain} 

\newcommand{\blind}{0}

\begin{document}

\def\spacingset#1{\renewcommand{\baselinestretch}%
{#1}\small\normalsize} \spacingset{1}


\if0\blind
{
  \title{\bf An optimal dynamic treatment regime estimator for indefinite-horizon survival outcomes}
\author{
  \textbf{Jane She} \\
  Department of Biostatistics, University of North Carolina\\
  jane.she@unc.edu  \\
  \and
  \textbf{Matthew Egberg} \\
  Department of Pediatrics, Division of Pediatric Gastroenterology, \\
  University of North Carolina School of Medicine, \\
  matthew.egberg@med.unc.edu \\
    \and
 \textbf{Michael R. Kosorok} \\
  Department of Biostatistics, University of North Carolina, \\
  kosorok@unc.edu
}
  \maketitle
} \fi

\if1\blind
{
  \bigskip
  \bigskip
  \bigskip
  \begin{center}
    {\LARGE\bf Title}
\end{center}
  \medskip
} \fi

\bigskip
\begin{abstract}
We propose a new method in indefinite-horizon settings for estimating optimal dynamic treatment regimes for time-to-event outcomes. This method allows patients to have different numbers of treatment stages and is constructed using generalized survival random forests to maximize mean survival time. We use summarized history and data pooling, preventing data from growing in dimension as a patient's decision points increase. The algorithm operates through model re-fitting, resulting in a single model optimized for all patients and all stages. We derive theoretical properties of the estimator such as consistency of the estimator and value function and characterize the number of refitting iterations needed. We also conduct a simulation study of patients with a flexible number of treatment stages to examine finite-sample performance of the estimator. Finally, we illustrate use of the algorithm using administrative insurance claims data for pediatric Crohn's disease patients.

\end{abstract}

\vspace{5mm}

\noindent%
{\it Some keywords:} Reinforcement Learning, Infinite Horizon, Indefinite Horizon, Survival Analysis, Random Forests, Machine Learning, Q-learning, Precision Medicine, Dynamic Treatment Regimes

\spacingset{1.45}
\section{Introduction}
\label{sec:intro}

\input{introduction.tex}

\section{Methods}
\label{sec:meth}
\subsection{Data Notation and Assumptions}

\input{notation.tex}

\subsection{Method Overview}

\input{method_overview.tex}

\section{Theoretical Properties}
\label{sec:theory}
\subsection{Consistency of the forest}

\input{consistency_forest.tex}

\subsection{Consistency of the value function}
\label{sec:value}

\input{consistency_value.tex}

\section{Simulations}
\label{sec:sims}

\input{simulations.tex}

\section{Example in Pediatric Crohn's Disease}
\label{sec:realdata}
\input{RDA.tex}

\section{Discussion}
\label{sec:disc}

\input{disc.tex}

\section{Competing interests}
No competing interest is declared.

\section{Author contributions statement}

J.S., M.E., M.R.K. conceived the ideas, analyzed the results, and wrote and reviewed the manuscript.

\section{Acknowledgements}
\label{sec:ack}

The authors would like to thank Xian Zhang for her extensive knowledge and advice about how to use administrative claims data.

\section{Data Availability}
Code reproducing the simulations can be found at github.com/js9gt/IHsurvrf with additional code for the data application portion at github.com/js9gt/RL\_IBS with additional README information for package use.

\section{Funding}
\label{sec:ack}

The first and last author were supported in part by the National Center for Advancing Translational Sciences (NCATS), National Institutes of Health, through Grant Award Number UM1 TR004406. The content is solely the responsibility of the authors and does not necessarily represent the official views of the NIH.

The last author's work for this paper was also supported in part by the Center for AI and Public Health at the University of North Carolina at Chapel Hill.

The project described was also supported in part by the National Center for Advancing Translational Sciences, National Institutes of Health, through Grant K12TR004416. The content is solely the responsibility of the authors and does not necessarily represent the official views of the NIH.

\section{Supplementary Material}

The Supplementary Material contains extra details on proofs, simulations, and the example data analysis.





\bibliographystyle{apalike} 
\bibliography{refs}

\newpage

\section*{\centering \Large Supplementary Material}

\input{Supplement.tex}

\end{document}

%% file: introduction.tex


In this paper, we propose a dynamic treatment regime (DTR) estimator in the indefinite-horizon setting for survival outcomes through non-parametric modeling. Our method is flexible, allowing patients to have differing number of treatment visits as well as differing visit lengths. The estimator operates by maximizing the truncated mean survival time. The estimator also has robust censoring assumptions, including the handling of dependent censoring. For an introduction and overview of DTRs, please refer to \cite{kosorok2019precision}.

On a high level, the algorithm uses recursive fitting to iteratively optimize outcomes over increasingly longer time trajectories. The process continues until reaching baseline, thereby optimizing outcomes across a patient's entire treatment trajectory. The primary outcome of interest is the truncated mean survival time. The recursive framework integrates nonparametric model fitting through generalized survival random forests to estimate remaining life at each visit. Rather than fitting multiple separate models which may compound in error, the final result is a single-model DTR estimator. This novel feature is particularly advantageous in the indefinite horizon setting and evaluates all patients across their entire trajectories using a single policy. \cite{clifton2020q} provides a general introduction to the recursive framework mentioned, which is applied to survival data in this context. 

Estimating a DTR is often of interest for chronic conditions such as diabetes, arthritis, mental illness, inflammatory bowel disease, etc., which require multiple treatments over time \newline \cite{chakraborty2014dynamic}. 
As such, DTRs use data-driven approaches to maximize long-term outcomes for patients, often employing machine learning methodologies. For example, in diabetes care, it may be of interest to estimate timing and dosage of insulin treatments, which are given frequently, with the overall goal of controlling a patient's glucose levels in the long-term, leading to reduced morbidities including kidney disease, vision deficits, and/or nerve damage.

Often, machine learning and reinforcement learning are employed in estimating DTRs, with a variety of established methods such as  Q-learning  \cite{watkins1992q} \cite{clifton2020q}, V-learning \cite{luckett2019estimating}, backwards and sequential outcome weighted learning \cite{zhao2015new}, and A-learning \cite{blatt2004learning, schulte2014q}. 

There are three common multi-visit decision settings considered in precision medicine, defined by the number of decision points and their characteristics: finite horizon, infinite horizon, and indefinite horizon. 
The methods presented in this paper are targeted towards the indefinite horizon setting where the number of decision points is large, but finite. Here, the challenge in estimating DTRs comes from the need for a flexible number of treatment visits for patients, as the number of decision points is not fixed and patients may receive differing numbers of treatment sequences. 

Q-learning has been employed in a variety of different reinforcement learning problems related to precision medicine such as in \cite{zhao2011reinforcement} \cite{moodie2012q}, \cite{song2015penalized} \cite{zhang2015using}, and has been utilized in all three time horizon settings. The algorithm relies on backwards recursion to maximize expected cumulative utility, given that future actions will follow the optimal decision strategy. However, in the case where there is patient drop out, leading to censored data, and a varying number of decision points per patient, standard backwards recursion becomes challenging as it is not clear what the final decision point is.

To address both censoring with survival outcomes as well as flexibility in the number of treatment visits, \cite{goldberg2012q} employed a novel Q-learning algorithm to estimate the optimal DTR in finite horizon settings to maximize the mean survival time. The authors construct an auxiliary Q-learning model by modifying the data so observations have the same number of visits, then use backwards recursion along with inverse-probability-of-censoring weighting (IPCW) to account for censored observations. However, the algorithm assumes that censoring is independent of event times and covariates, which may be restrictive. 

In a finite horizon setting with survival outcomes, \cite{simoneau2020estimating} propose a DTR estimator by solving weighted ordinary least squares using backwards recursion and IPCW to account for right-censoring. The resulting estimator is doubly robust, and uses the mean survival time for the outcome. While this method allows for a flexible number of treatment visits and censoring, the resulting estimator is limited to a binary treatment class. Additionally, the method uses the accelerated failure time (AFT) model, which may lead to model misspecification. 

More recently, \cite{cho2023multi} proposed a generalized survival random forest utilizing a Q-learning-based algorithm along with survival random forests by \cite{ishwaran2008random} to estimate the optimal DTR in finite horizon settings. This method considers both censoring in survival outcomes and flexibility in number of treatment visits while relaxing the censoring assumptions to allow for dependent censoring. 
The method allows for censoring to be conditionally independent given the covariates, prior history, and prior treatments, and employs non-parametric modeling. The survival random forests have desirable theoretical properties including consistency of estimated regime values as well as desirable convergence rates. However, such a method is designed for the finite-horizon setting and cannot be reliably used for a larger number of visits, as the method fits separate models for each visit, compounding in error if applied to a large number of visits.

In the indefinite horizon setting, \cite{ertefaie2014constructing} develops a DTR estimator when there is no fixed last visit which operates through estimating equations and does not require backwards induction. 
However, this method does not consider survival outcomes or censoring. An additional limitation is that the method is only valid in cases where decision points have fixed spacing, limiting its applicability when treatment frequency may depend on patient health.

While estimation of multi-visit regimes for survival data is a relatively new area, there have been a variety of methods developed to estimate DTRs for survival outcomes, such as \cite{jiang2017estimation}, \cite{huang2014optimization} and \cite{goldberg2012q}, \cite{simoneau2020estimating}, \cite{cho2023multi} which were previously discussed. A few methods are available for the indefinite horizon problem such as \cite{zhou2024estimating},  \cite{luckett2019estimating}, and \cite{ertefaie2014constructing}, however, as far as we know, there are none applicable to survival outcomes.

To our knowledge, there have been no existing methods available for DTR estimation in the indefinite horizon case with survival outcomes and right censored data while allowing for a flexible number of treatment visits. We develop a novel DTR estimator for censored survival outcomes allowing for a flexible number of treatment visits and dependent censoring. Our estimation operates through maximization of the mean survival time for a flexible number of treatment arms. By adapting the framework laid out in \cite{cho2023multi}, we non-parametrically model the conditional survival probability using multiple iterations of model fitting with the generalized survival random forest, pooling data across multiple time points. Dependent censoring is allowed, but is conditionally independent given the history and covariates. The key contribution of this work lies in the applicability to the indefinite horizon scenario, while maintaining flexibility in model assumptions.


\vspace{-7mm}

%% file: notation.tex

The data consists of $n$ independent patients with a maximum of $\tilde{K}$ visits, though this individual number of visits, $K_i \leq \tilde{K}$, can vary by patient, where $\tilde{K}$ is a constant. The overall study length for all patients is $\tau > 0$, where $\tau$ is finite. There are a finite number of possible treatments and treatment space is represented as $\mathcal{A}$. At each visit $k$, the $i$th patient, if still at risk, will receive treatment $a_{k,i} \in \mathcal{A}$, for $k = 1, ... K_i$. 

At each visit, a patient will be followed until they experience failure, advancement to the next visit, or censoring. From the beginning of visit $k$, these times can be denoted as $T_{k,i}$, $U_{k,i}$, and $C_{k,i}$, which are dependent on the historical information available at the beginning of the visit. The observed length of the visit, $X_{k, i}$ is the minimum of these three quantities. Uncensored visit length is based on the minimum of failure and advancement to the next visit, defined as $V_{k, i} = T_{k, i} \wedge U_{k, i}$. The visit's censoring indicator is denoted as $\delta_{k, i} = I(V_{k, i} \leq C_{k, i})$. The failure indicator is denoted as $\gamma_{k, i} = I(T_{k, i} \leq U_{k, i})$, which identifies whether a patient failed (1) or moved onto the next treatment visit (0). The cumulative time from baseline at visit $k$ is defined as $B_{k, i} = \sum_{j = 1}^{k - 1}X_{j, i}$, where $k >1$ and $B_{1, i} = 0$. At each visit, some covariate information becomes newly available, and is represented by $Z_{k, i} \in \mathcal{Z}_k$. $T$ denotes the overall failure time across all visits.

At each visit $k$, we use a summary of the observed history, rather than the whole history. We denote this using $H_{k, i}$, which contains a summary of past covariates such as baseline time ($B_{k', i}$), the total number of each treatment the patient has had in the past, the number of previous visits, the previous visit's length $X_{k-1}$, baseline variables, and newly available covariate information $Z_{k,i}$, where $k' = 1, 2, ... k -1$. We require that the dimension and variable definitions of $H_{k, i}$ remain the same across visits, $k$, though their values can vary. we denote this domain as $\mathcal{H}$; i.e. $H_{k, i} \in \mathcal{H}$ for $k = 1, ... , K_i$, and $i = 1, ... n$.

\begin{figure}[H]
    \centering
    \includegraphics[width=0.8\linewidth]{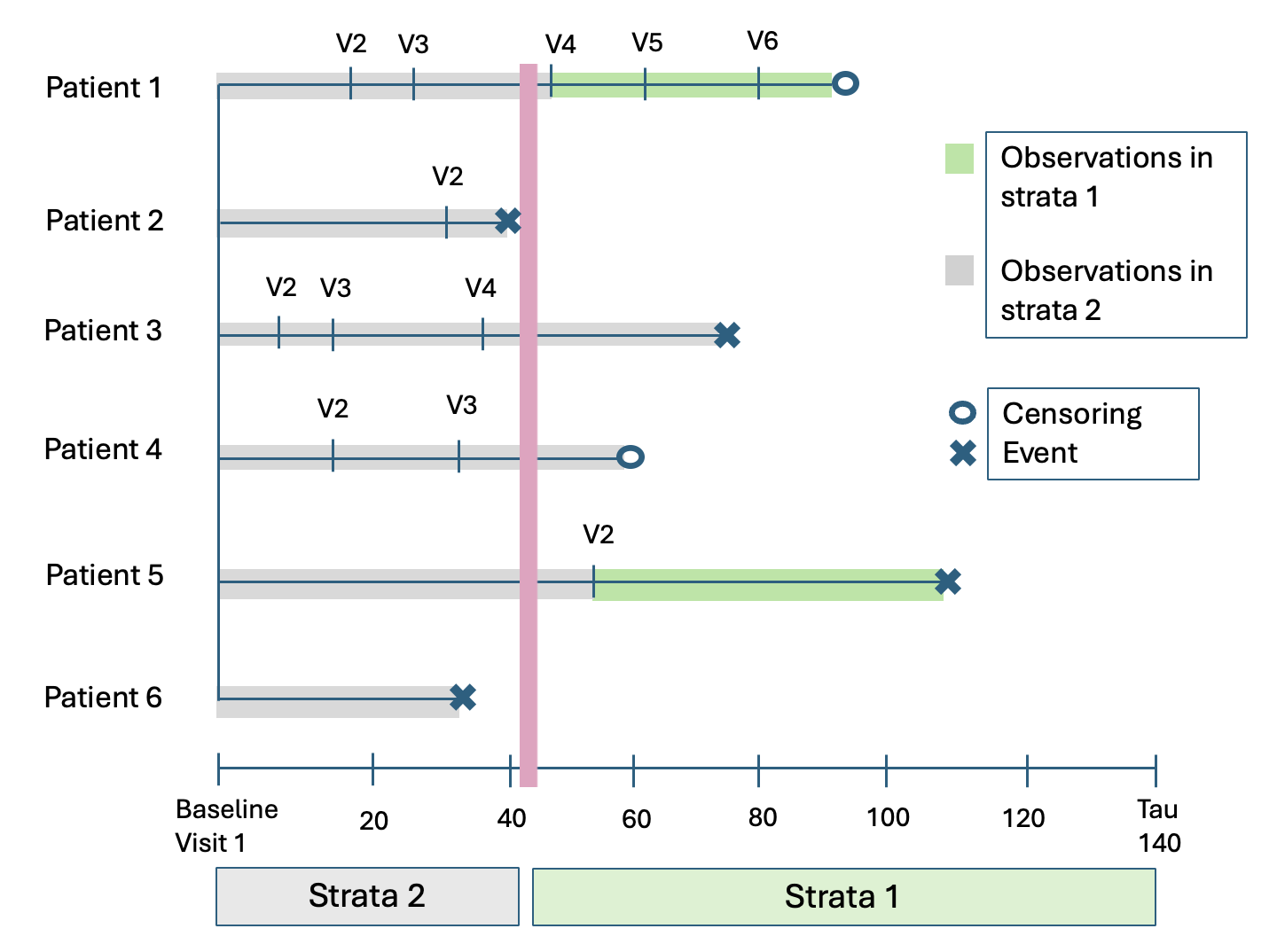}
    \caption{A 2-strata example of 6 patients, where V1, V2... VK refer to visits}
    \label{fig:strata_visual}
\end{figure}

Our goal is to seek optimal treatment assignments for each visit. For flexibility, we allow the  division of data into multiple strata based on the cumulative time from baseline. 
We will denote each strata as $l = 1, ... W$.
The algorithm will employ a finite number of generalized survival random forests, one for each strata, where the final estimator combines strata information to fit a single model overall. A visualization of the $2$ strata example can be seen in Figure \ref{fig:strata_visual}, with more discussion to follow. Additionally, a basic table summary of the notation can be found in the supplement. 

The algorithm also works in the single strata case, where a single generalized survival random forest is fit. The purpose of the multi-strata approach is dependent on the data, where larger divides between short and long trajectory patients may lead to differing characteristics, so it may be of interest to initialize different models.

The dynamic treatment regime (DTR) of interest is a function, $\pi$, which maps from patient historical information at $k$ to an action at $k$; i.e. $h_{k,i} \in \mathcal{H} \rightarrow a_{k,i} \in \mathcal{A}$. This is a single policy applied across all visits and patients based on patient history, and visit times. Although the DTR, $\pi$, is fixed for all patients and visits, the actual sequence of treatments can vary, since patient history will vary. The optimal dynamic treatment regime, denoted by $\pi^*$ is the sequence of treatments optimizing an overall criterion of $\phi$, for all patients in the population.

$\phi$ is the mean survival time truncated by the study length $\tau$ represented as $\phi(S) = -\int_{t>0} (t \wedge \tau) \, dS(t)$. The goal is to maximize mean survival time up to $\tau$.

%% file: method_overview.tex
As in the case of \cite{cho2023multi}, the method is based on the remaining life, $L_k$ at visit $k$ being defined as a convolution of $L_{k+1}$ and the uncensored visit length $V_k$. Under the covariate-conditionally independent censoring assumption, the censored cases have their probabilities redistributed to allow for optimizing the distribution of $L_k$, which is assumed to be independent of $C_k$ conditional on $H_k$ and $A_k$, for $k = 1, 2..., K_i$.
For ease of notation and to avoid confusion, we drop the $i$ index for $K_i$ from now on in favor of using $K$, and differentiate this from the constant $\tilde{K}$, which represents the maximum visit in the study.

$S_k(t | H_k, A_k = a)$ represents the survival curve at the $k$th visit based on patient history and treatment, and are temporary visit-wise estimates used in the algorithm, until we optimize over the entire patient trajectory, which will be elaborated further. As previously specified, $H_k$ includes previous visit length, baseline time, and other covariates. The dynamic treatment regime (DTR) is estimated using a single random forest model, which combines the multi-forest procedure based on the number of prespecified strata divisions. The end goal is to use backwards recursion to optimize $S_1(H_1, A_1 = a)$ when all future visits are optimal. The estimated DTR can then be applied to all visits across all strata. Assuming the causal inference assumptions of stable unit treatment value assumption (SUTVA), sequential ignorability, and positivity, $S_k(t | H_k, A_k = a)$ is rewritten into the counterfactual quantity of $S_k^a(t | H_k)$. The causal assumptions are defined in greater detail in section \ref{sec:theory}. However, our final curve of interest is $S_1(t | H_k, A_k = a)$. 

Each forest is fit by pooling data together within the strata, treating each patient's visit as a separate observation. The algorithm for a single strata follows the framework outlined below, with more details later about specific inputs and outputs, and how this can be extended to a multi-strata approach.

\underline{Framework Overview}

\begin{enumerate}
    \item Fit a pooled survival random forest (initialization), resulting in survival curves for each observation across all relevant visits.
    
    \item Using step 1: predict optimized survival curves for the final visit, $k = K$, and move backwards to the previous visit and append $k - 1$ information. 

    \item Refit forest using step 2's augmented survival curves, resulting in survival curves which are now optimized over both visits $k$ and $k - 1$.
    
    \item Repeat steps 2-3, for $k = K - 2, ..., 1$: refit forest until $\#$ of iterations = $\tilde{K}$ using backwards recursion. Eventually, survival curves will be maximized over the entire patient trajectory, resulting in a DTR estimator at baseline.
\end{enumerate}


For the initial pooled forest, we select the treatment maximizing the criterion for each visit without accounting for long-term treatment effects. However, in general, we know visit-wise optimization may not result in the optimal DTR.

A modified kaplan-meier estimator defined in \cite{cho2023multi} was used for tree partitioning and estimation of survival probabilities in the terminal node of each forest where the number of trees and minimum terminal node size are hyperparameters. A generalized log-rank test was used for the splitting rules, which was first developed in \cite{cho2022interval}. Refer to the B.2 supplementary section in the original paper for details about splitting rules and node partitioning.

After dividing data observations into strata, each strata will have their own generalized survival random forest fit for available observations. This is because patients who survive into later strata may fundamentally differ from patients in the earlier strata. However, earlier strata will use optimized information from later strata, so the resulting estimator relies only on a single random forest model. Note that the visits are numbered $k = 1, ... \tilde{K}$, while the strata are numbered $l = W, ..., 1$.
A visual representation has been provided in Figure \ref{fig:strata_visual}. 

Using backwards recursion, we begin with strata $l =1$ and pool observations within the strata to fit a generalized survival random forest. This pooling considers each visit, $k$, as a separate observation, and optimizes the survival distribution for each visit. It should be noted, that the following estimates are only applicable to observations within the strata. 
The result is a set of intermediate optimal actions $\hat{\pi}_{j=1} = argmax_{a \in \mathcal{A}} \phi\{\hat{S}_k(. |H_k, A_k = a)\}$ and an estimated optimal set of survival probabilities $\hat{S}_k^*(. |H_k, A_k = \hat{\pi}_{j=1}(H_k))$, for all $k$ within strata $l$, where $j$ is an iteration counter for each model refitting, which will estimate a new set of optimized treatment rules and survival  curves each time. \underline{Details can be found in step 1 of algorithm 1.} Here, the series of intermediate optimal actions is not the final DTR, which will be specified in more detail.











The result of this step 1 is a trained random forest. We can initialize the forest for each strata independently, however, steps 2-4 must be conducted in order of strata, where the first strata must finish steps 2-4 optimization before moving onto the second strata, and so forth, to ensure all future visits are optimal and that we have maximized over the entire time period of the strata.

\begin{figure}[h]
    \centering\includegraphics[width=1.1\linewidth]{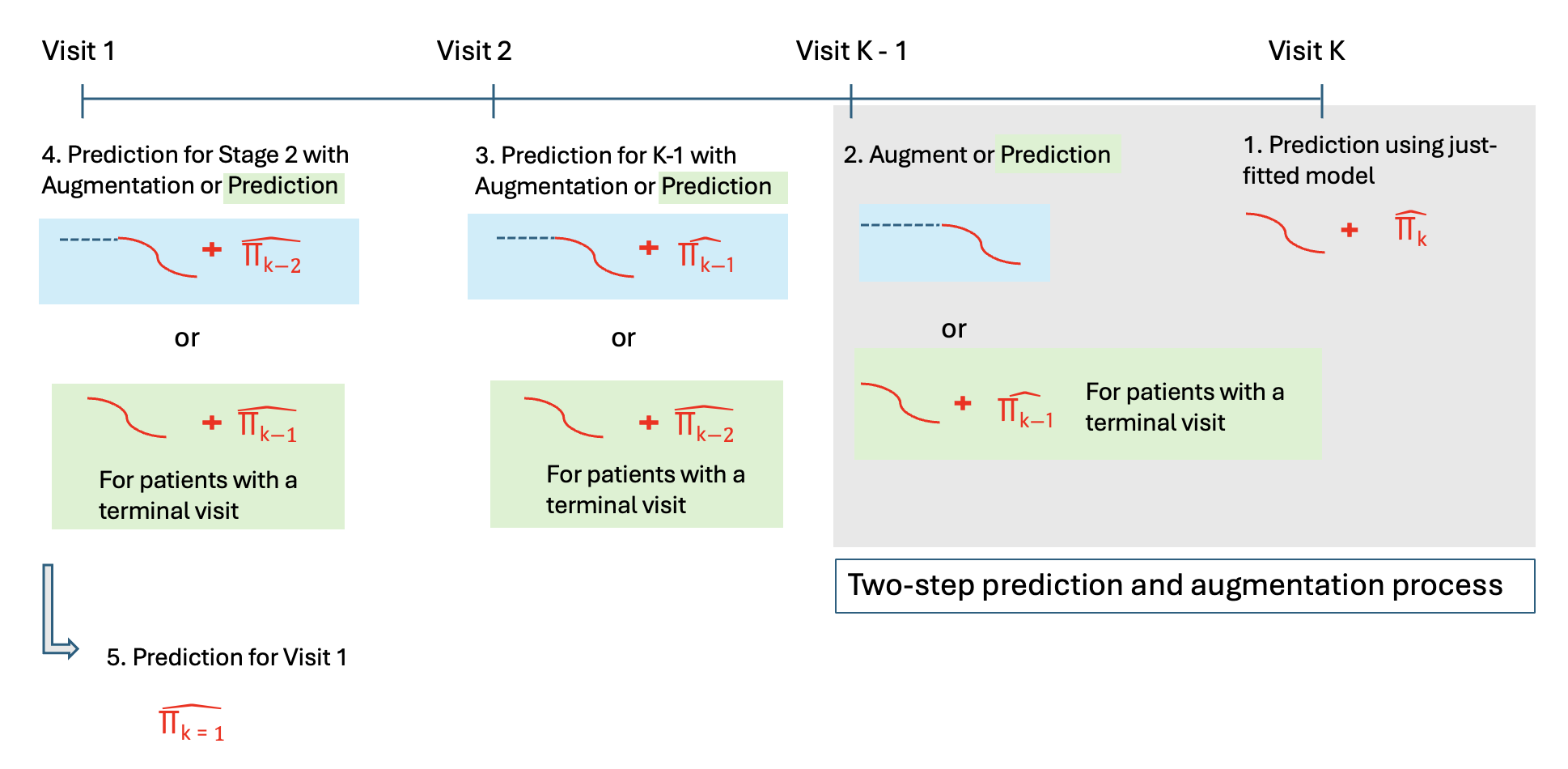}
    \caption{Visual illustration of prediction process for refitting using $\tilde{K} = 4$ visits in one strata}
    \label{fig:strat1_pred}
\end{figure}

Each refitting iteration employs a probabilistic augmentation where the optimized survival information of the next visit is carried  back to the previous visit. Cases when the next visit belongs to a different strata is discussed later, but uses the same principles. Mathematically, at visit $k = K - 1$, the optimal survival information at the next visit can be represented as $\hat{S}^*_{K+1}$; the $i$th patient, which has a visit length of $X_k$ then has an augmentation so that the survival probability of the remaining life is $\hat{S}^*_{K}(t - X_{K,i} | H_{K, i})$. For those who have already experienced failure or censoring by the end of the visit, no augmentation is required. Again, this process is only done for observations within the current strata $l$. For a more detailed description of augmentation, please refer to section 2.2 in \cite{cho2023multi}. A visual example of the two-step augmentation process is provided in Figure \ref{fig:strat1_pred}. \underline{Details are in step 2 of algorithm 1.}


Finally, the first refitting iteration of strata $l =1$'s forest uses the predicted intermediate treatments and augmented optimal survival information as inputs to estimate another set of intermediate optimal actions and survival curves, using observations belonging to strata $l$. Again, these intermediate estimates of actions do not yet reflect the final estimated DTR, as we do not yet consider all future visits optimal from baseline. Now, the survival curves are optimized over two visits, going backwards. \underline{Details can be found in step 3 of algorithm 1.} 

As we cycle through each iteration, we optimize over a longer span of visits, with each successive iteration, $j$, optimizing over $j$ visits. The final refitting iteration optimizes over the entire patient trajectory in the strata. We then move backwards to the earlier strata, $l + 1$, and repeat the same process. 

Note that the prediction process relies on using the later optimized strata's survival information for appending. Meaning, if a patient has visit $k$ in strata $l+1$, but visit $k + 1$ in strata $l$, visit $k + 1$'s optimized information from the strata $l$ forest should be carried back. The optimal actions from the final refitting iteration in the earliest strata is the estimated DTR, which adapts to the history and relative position to the overall $\tau$.


\newpage 

\newpage

\input{algorithm}

%% file: algorithm.tex
\setcounter{algorithm}{0}  

\begin{algorithm}[H]
  \label{alg:strata1_DTR}
  
\small
  \caption{Proposed DTR Estimator for A Finite Number of Strata}

  \textbf{Result:} A dynamic treatment regime estimate $\hat{\pi}$.

\underline{\textbf{Step 1: First Pooled Survival Random Forest}}

    \textbf{Input:} $\{(H_k, A_k, \delta_k, V_k)\}^{n_k}_{i=1}$ for eligible $k = 1, 2...K$ in strata $l$.
  
\begin{itemize}
\vspace{-1.5mm}
  
    \item Obtain $\hat{S}_k(\cdot | H_k, A_k)$ and $\hat{\pi}_{j = 1}(H) = \arg\max_{a \in A} \phi\{ \hat{S}_k(\cdot | H_k = H, A_k = a) \}$ via the generalized random survival forest \cite{cho2023multi}.

    \vspace{-1.5mm}
    \item Define $\hat{S}_k^\ast(\cdot | H_k) = \hat{S}_k(\cdot | H_k, A_k = \hat{\pi}_{j = 1}(H_k))$.

    \vspace{-1.5mm}

  \end{itemize}

  \underline{\textbf{Step 2: Convergence Predictions}} 
  \begin{enumerate}
      \item \textbf{For visit $k = K$, Input:} $\{(H_k, A_k = \hat\pi_{j =1}(H), \delta_k, V_k)\}^{n_k}_{i=1}$ for eligible $k =K$ in strata $l$ into Step 1's forest.

      \vspace{-2.5mm}

  \begin{itemize}
  
    \item Obtain $\hat{S}_k(\cdot | H_k, A_k)$ and update $\hat{\pi}_{j = 1}(H) = \arg\max_{a \in A} \phi\{ \hat{S}_k(\cdot | H_k = H, A_k = a) \}$.
    \vspace{-1.5mm}
    
    \item Define $\hat{S}_k^\ast(\cdot | H_k) = \hat{S}_k(\cdot | H_k, A_k = \hat{\pi}_{j = 1}(H_k))$.

\vspace{-1.5mm}
  \end{itemize}

  \item \textbf{For visit $k = K-1$}:

  \vspace{-2.5mm}

   \begin{itemize}

    \item Perform stochastic augmentation: $S_{k, i} = \hat{S}^\ast_{k+1}(t - X_{k, i} | H_{k+1} = h_{k+1, i})$ with $S_{k, i}(t) = I(t \leq X_{k, i})$ if $k = K-1$ or $\delta_{k, i} = 0$, choice of $\hat{S}^*_k$ based on visit $k = K$ strata membership.

\vspace{-1.5mm}
    \item \textbf{If $\delta_{k, i} = 1$ or $\gamma_{k, i} = 1$, Input:} $\{(H_k, A_k, \delta_k, V_k)\}^{n_k}_{i=1}$ for eligible $k = K-1$ in strata $l$ into Step 1's forest.

\vspace{-1.5mm}
    \begin{itemize}
            \item Obtain $\hat{S}_k(\cdot | H_k, A_k)$ and $\hat{\pi}_{j = 1}(H) = \arg\max_{a \in A} \phi\{ \hat{S}_k(\cdot | H_k = H, A_k = a) \}$.

    \item Define $\hat{S}_k^\ast(\cdot | H_k) = \hat{S}_k(\cdot | H_k, A_k = \hat{\pi}_{j = 1}(H_k))$.

    \vspace{-1.5mm}
    \end{itemize}

  \end{itemize}

  \item \textbf{Repeat for pairs \{$k = K-1$ in (i), $k = K-2$ in (ii)\}.... \{$k = 2$ in (i), $k = 1$ in (ii)\} }
\vspace{-1.5mm}

  \end{enumerate}

  \underline{\textbf{Step 3: Re-fitting Forest}}

      \textbf{Input:} $\{(H_k, \hat{\pi}_k, \delta_k, \hat{S}^\ast_{k+1}(\cdot - X_{k} | H_{k+1}))\}^{n_k}_{i=1}$ for eligible $k = 1, 2, ..., K-1$ in strata $l$ and $\{(H_k, \hat{\pi}_k, \delta_k, \hat{S}^\ast_{k}(\cdot - X_{k} | H_{k}))\}^{n_k}_{i=1}$ for eligible $k = K$ in strata $l$.

\vspace{-1.5mm}
      \begin{itemize}
            \item Obtain $\hat{S}_k(\cdot | H_k, A_k)$ and $\hat{\pi}_{j = 2}(H) = \arg\max_{a \in A} \phi\{ \hat{S}_k(\cdot | H_k = H, A_k = a) \}$.

\vspace{-1.5mm}
    
    \item Define $\hat{S}_k^\ast(\cdot | H_k) = \hat{S}_k(\cdot | H_k, A_k = \hat{\pi}_{j= 2}(H_k))$.
    \end{itemize}

    \vspace{-1.5mm}

    \underline{\textbf{Step 4:}} Repeat steps 2-3 until total number of iterations $j$ = $\sum_l M_l$ ($M_l$ is maximum number of visits in a strata), for each strata. Finally, obtain $S_1^*(\cdot | H_1, A_1)$ and $\hat{\pi}$, where the former represents the survival information at baseline optimized over all future visits, and the latter is the DTR.  \\
 
\vspace{-1.5mm}
\end{algorithm}

%% file: consistency_forest.tex






To establish uniform consistency of our forest, we utilize the following assumptions for terminal node sizes and splitting rules.

\vspace{2mm}

\setlength{\parindent}{0pt}

\textbf{Assumption 1 (Terminal node size, polynomial).} The minimum size $n_{\text{min}}$ of the terminal nodes grows at the rate $n_{\text{min}} \asymp n^\beta$, $\frac{1}{2} < \beta < 1$, where $a \asymp b$ implies that both $a = \mathcal{O}(b)$ and $b = \mathcal{O}(a)$.

\vspace{2mm}

\textbf{Definition 1 (Random-split and $\alpha$-regular trees and forests).} A tree is called a random-split tree if each feature is given a minimum probability $\left(\frac{\varphi}{d}\right)$ of being the split variable at each intermediate node, where $0 < \varphi < 1$ and $d$ is the dimension of the feature space. A tree is $\alpha$-regular if every daughter node has at least an $\alpha$ fraction of the training sample in the corresponding parent node. A random forest is called a random-split ($\alpha$-regular) forest if each member tree is random-split ($\alpha$-regular).

\vspace{2mm}

\textbf{Assumption 2 ($\alpha$-regular and random-split trees).} Trees are $\alpha$-regular and random-split with a constant $0 < \varphi < 1$.

\vspace{2mm}

\textbf{Assumption 3 (Lipschitz continuous and bounded survival and censoring probability).} For $k$ within strata $l$ and for the $j$th refitting iteration, there exist constants $L_S$ and $L_G$ such that $|S_k(t \mid h_1) - S_k(t \mid h_2)| \leq L_S \| h_1 - h_2 \|$ and $|G_k(t \mid h_1) - G_k(t \mid h_2)| \leq L_G \| h_1 - h_2 \|$ for all $h_1, h_2 \in \mathcal{H}$, $t \in [0, \tau_{lj} - B_k]$ and $k = 1, 2, \dots, K$, where $G_k$ is the CDF of the censoring survival distribution at the $k$th visit and $S_k(\tau^{-}_{lj} - B_k \mid h_k) G_k(\tau^{-}_{lj} - B_k \mid h_k) > c_1$ for all $h_k$ with $\gamma_{k-1} = 0$, i.e., patients who made visit $k$, and some constant $c_1 > 0$. 

\vspace{2mm}

\textbf{Assumption 4 (Weakly dependent historical information).} The patient history information \( H_k \) at visit \( k \) is given as a \( d \)-dimensional vector lying in a subset \( \mathcal{H} \) of \( [0, 1]^{d} \), which contains relevant information about visit $k$, including visit-specific time and time-relative to $\tau$. The baseline distribution of patient histories, $H_1$, has support $\mathcal{W}_1$, where $\mathcal{W}_1 = \{h \in \mathcal{H}_1: Pr(H_1 = h) > 0 \}$. The transition distribution, denoted as $Pr(H_2 = h' | H_1 = h, A_1 = a)$ has support $\mathcal{W}_2$, where $\mathcal{W}_2 = \{h' \in \mathcal{H}_2 : Pr(H_2 = h' | H_1 = h, A_1 = a) 
> 0, \text{ for all } h \in \mathcal{W}_1\}$. The baseline and transition distributions are bounded so that 
\[
\frac{1}{\zeta} \leq f_{H_1}(h) \leq \zeta \quad \text{and} \quad \frac{1}{\zeta} \leq f_{H_{2}}(h', a) \leq \zeta
\]
for all \( h \in H_1 \) and \( h' \in H_2 \), and $a \in \mathcal{A}$, for some constant \( \zeta \geq 1 \). We allow a subset of $\mathcal{H}$ to have finite support. This assumption allows for non-hyper-rectangular or categorical history spaces.

\vspace{2mm}

We first prove uniform consistency of our initial pooled random forest via Theorem 1, which closely follows the proof of Theorem 2 of \cite{zhou2024optimal}. Next, we prove uniform consistency of each re-fitting iteration's forest via theorem 2. The proof will closely follow the proof in \cite{cho2023multi}, but has been expanded upon for clarity. We should note that the proof of Theorem 2 relies on the legitimacy of backwards recursion, which is proved in the supplement. Finally, we will prove that we have a finite number of iterations in Lemma 1. 

\vspace{2mm}

\textbf{Assumption 5 (Boundedness of visits).} We have a maximum visit length, $max(\mathcal{X}_l)$ for each strata $l$, and a minimum visit length in each strata, $min(\mathcal{X}_l) > 0$, where $P(\mathcal{X}_l > max[\mathcal{X}_l]) = 0 $ and $P(\mathcal{X}_l \geq min[\mathcal{X}_l]) = 1$. In our setting, the visit length must be at least one day, so in our real data analysis, $min[\mathcal{X}_l] \geq 1$. This will be used in our proof that the number of iterations, $j$ is finite for Lemma 1.

\vspace{2mm}

\textbf{Assumption 8 (Stable unit treatment value assumption).} Each person's counterfactural dynamics, such as failure time, time to next treatment, and censoring, are not affected by the treatments or histories of other patients. Moreover, each treatment has only one version. 

\vspace{2mm}

\textbf{Assumption 9 (Sequential ignorability).} Treatment assignment is given independently of the counterfactual outcomes, conditionally on the individual's historical information, where the outcomes include all future random quantities such as failure time, time to the next treatment, and censoring time. 

\vspace{2mm}

\textbf{Assumption 10 (Positivity).} At each $k = 1, 2, ..., \tilde{K}$, given historical information, the probability of having each treatment $a \in \mathcal{A}$ is greater than a constant $L > 0$, or $pr(A_k = a | H_k = h) > L$ for all $a \in \mathcal{A}, h$ such that $pr(H_k = h) > 0$.

\vspace{2mm}

\textbf{Lemma 1 (Finite re-fitting iterations).} Suppose that assumption 4 holds, then the number of iterations, $j$ for each strata, where $j = 1... J$ is finite, and is equal to the sum of all the maximum number of visits within a strata, denoted as $M_l$. The proof along with a visual illustration are included in the supplement.

\vspace{2mm}

{\setlength{\parindent}{15pt}

\textbf{Theorem 1 (Uniform consistency of pooled random survival forest).} Suppose that Assumptions 1–4 hold. Let $\hat{S}_k$ be an estimator of visit $k$ survival probability $S_k$ that is uniformly consistent in both $h \in \mathcal{H}$ and $t \in [0, \tau_{lj=1}]$ for all visits $k \in \mathcal{M}_l$. We define $\tau_{lj=1}$ as the length of time we maximize over for the first iteration, $j = 1$ for strata $l$. Then, the generalized random survival forest $\hat{S}_k(t \mid h)$ built based on $\{[H_{k,i}, \delta_{k,i}, X_{k, i}, A_{k,i}]\}_{i=1}^{\mathcal{M}_l}$ is uniformly consistent. Specifically, for each $k = 1, 2, \dots, K \in \mathcal{M}_l$, where $\tau_{lj=1} \leq \tau$, 

\vspace{2mm}
$\sup_{t \in [0, \tau_{1j=1}], h_k \in \mathcal{H}} |\hat{S}_k (t \mid h_k) - S_k (t \mid h_k)| \to 0 \text{ in probability as $n \to \infty$. } $


\vspace{5mm}

\textbf{Theorem 2 (Uniform consistency of optimized survival curves).} Suppose that Assumptions 1–5 hold. Let $\hat{S}_{k+1}$ be an estimator of the next-visit survival probability $S_{k+1}$ that is uniformly consistent in both $h \in \mathcal{H}$ and $t \in [0, \tau_{lj=1}]$ for all visits $k \in \mathcal{M}_l$, as proven from Theorem 1. Then, the generalized random survival forest $\hat{S}_k(t \mid h)$ built based on $\{[H_{k,i}, \delta_{k,i}, \hat{S}_{k+1}(\cdot - X_k \mid H_{k+1,i})]\}_{i=1}^{\mathcal{M}_l}$ is uniformly consistent. Specifically, for each $k = 1, 2, \dots, K \in \mathcal{M}_l$, where $\tau_{lj} \leq \tau$, and for each iteration $j = 2, ..., J$, and each strata $l$,

\vspace{2mm}
$\sup_{t \in [0, \tau_{lj}], h_k \in \mathcal{H}} |\hat{S}_k (t \mid h_k) - S_k (t \mid h_k)| \to 0 \text{ in probability as $n \to \infty$}$.

\vspace{2mm}

Each iteration is uniformly consistent only over the value of $\tau_{lj}$, and that $\tau_{lj=1} < \tau_{lj=2} < ... < \tau_{lj=J} \leq \tau$, with the number of iterations, J, being finite from Lemma 1. At baseline, $\hat{S}_1$ is consistent over $[0,\tau]$. The proof of theorem 2 is also in the supplement.

While we show consistency of each $\hat{S}_k$, we reiterate, that the final estimate of interest is the baseline survival curve when all future visits have already been optimized, denoted as $\hat{S}_1$. This will cover all visits and all strata.

%% file: consistency_value.tex
 Now, we show that the estimated DTR has a value consistent for the value of the true optimal regime, where the value, $\mathcal{V}$, is the criterion value of the survival probability if all individuals follow the treatment regime. This requires assumptions using the causal inference framework from \cite{rubin2005causal, Hernan2024-HERCIW}. We define a counterfactual outcome as the outcome of a patient if, counter to fact, they received a treatment option different from the actual treatment. We denote the counterfactual survival probability as $S_k^\pi(t | H_k)$, which denotes the survival probability of remaining life at visit $k$ if treatment was given according to a treatment rule, $\pi$. 

\vspace{2mm}

\setlength{\parindent}{0pt}

\vspace{2mm}

\textbf{Theorem 3 (Consistency of the dynamic treatment regime estimator).}  Let $\hat\pi$ denote the optimal dynamic treatment regime estimator built following algorithm 1, and let assumptions 1-2 (terminal node size; $\alpha$-regular and random-split trees) hold. Furthermore, assume assumptions 3-4 (Lipschitz continuous; weakly dependent historical information) and assumptions 6-8 hold (SUTVA; sequential ignorability, positivity). Then, 
$\bigg| \mathcal{V}(\pi) - \mathcal{V}(\hat\pi)\bigg| \rightarrow 0 \text{ in probability as $n \rightarrow \infty$.}$ The proof of Theorem 3 is in the supplement.

\vspace{-5mm}

%% file: simulations.tex
{\setlength{\parindent}{15pt}

We conducted a simulation study to evaluate the performance of the proposed estimator in comparison to the observed policy. The data generating mechanism follows an exponential distribution, where patients can have up to $\tilde{K} = 10$ visits during the hypothetical study, with a maximum trial length of $\tau = 1000$. We also test the setting where patients can have up to $\tilde{K} = 15$ visits, with a maximum trial length of $\tau = 1500$. At each visit, patients are given either treatment $A = 1$ or $A = 0$, with treatment allocation based on a bernoulli trial with propensity $\pi(H_{k}) = {[1 + \exp(-g(S_{1k}, S_{2k})^T\beta_{\pi})]}^{-1}$, 
where treatment propensity is based on two patient state variables (not to be confused for survival) at each visit $(S_{1k}, S_{2k})$. Mimicking the set-up in \cite{cho2023multi}, the coefficients for $\beta_{\pi}$ take different values for the observational data setting, where the observational setting uses $\beta_{\pi} = (0, -\frac{1}{2}, -\frac{1}{2})^T$.

At the initial visit, each patient has two random state variables generated, $S_{1k}, S_{2k} \sim U(0, 1)$. Subsequent states are dependent on the value of the prior state. For more details, please refer to the supplement. In addition, each patient has two baseline variables generated as part of the history, one from a $U(0, 1)$ distribution, the other from a binomial distribution with equal probability. At each visit, patient failure time ($T_k$), time to next visit ($U_k$), and censoring time ($C_k$) are generated from an exponential distribution using the hazard rates using a function of a subset of the history variables and covariates:

\vspace{-6mm}

\begin{align*}
\lambda_{T_k} &= \exp(g(Z_k, H_k)^T\beta_{T_k}) \\
\lambda_{U_k} &= \exp(g(Z_k, H_k)^T\beta_{U_k}) \\
\lambda_{C_k} &= \exp(g(Z_k, H_k)^T\beta_{C_k}),
\end{align*}

where $Z_k$ consists of the following covariates: two patient state variables which are newly updated ($S_{1k}, S_{2k}$) and the action taken in the current visit ($A_k$). The history variables included are the number of prior visits ($G_k$), cumulative length from baseline ($B_k$), prior visit length ($X_{k-1})$, and the interaction between all these terms. An intercept term is included as well. The coefficients in the $\beta$s are provided in the supplementary material.

Simulations were run with N = 300 and N = 500, with 200 simulation replicates each. The splitpoint between strata 1 and strata 2 was defined as the timepoint at which the minimum percentage of events in strata 2 was at least $30\%$. 
In the setting with 15 visits, we only considered the sample size of 500, due to computational intensity and time required to run with more visits. 
Performance of the proposed algorithm was compared to the observed policy, which generate treatment assignments based on a bernoulli trial with the aforementioned propensity scores. The value of the observed regime was estimated using a sample size of $10,000$ patients who were not subject to censoring, and applying the value function.

\vspace{-10mm}

\begin{figure}[h]
    \centering
    \includegraphics[width=\linewidth]{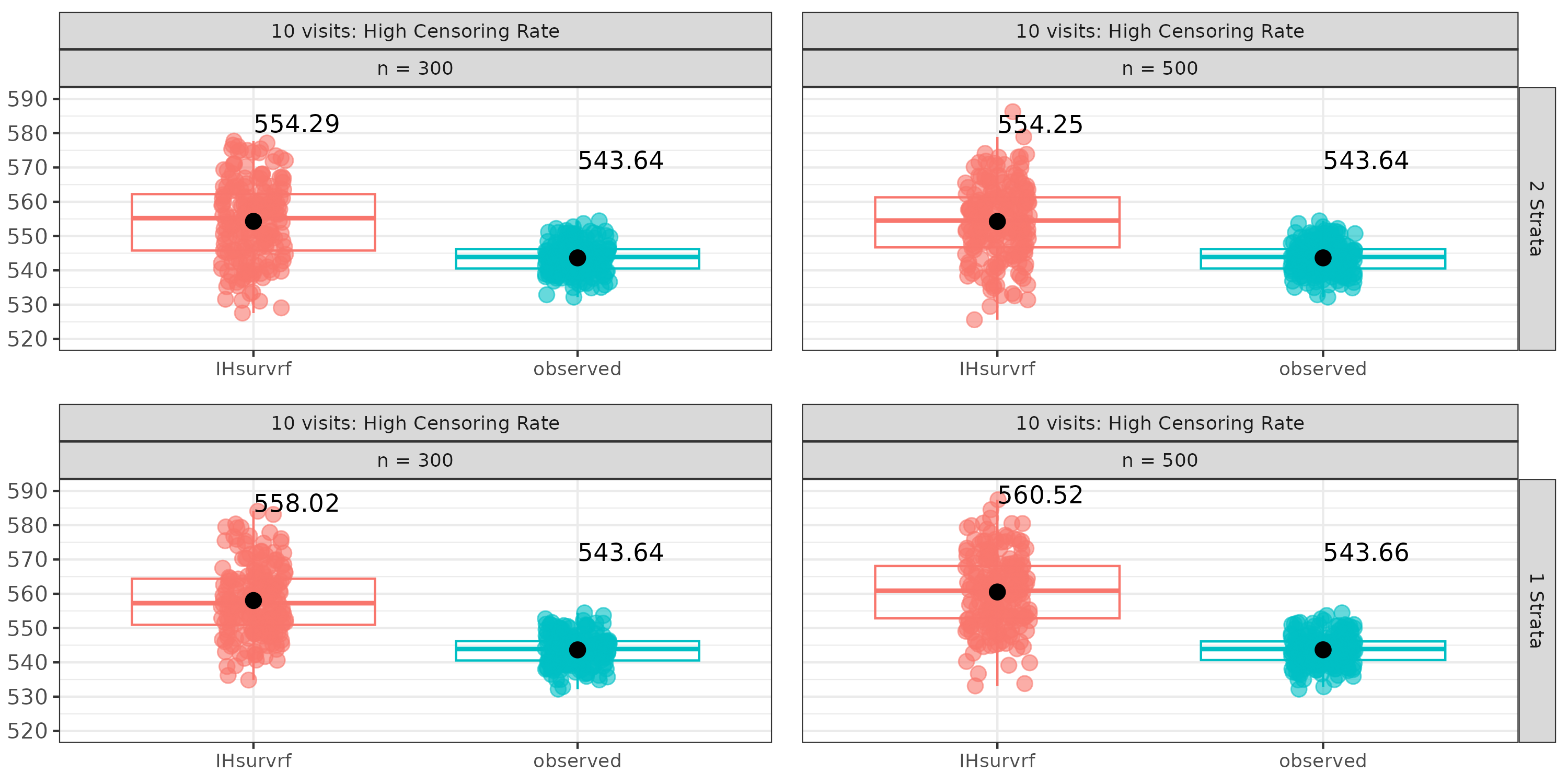}
    \caption{Boxplots for high censoring ($\sim$ 44\%) for 10 visits}
    \label{fig:enter-label}
\end{figure}

\begin{figure}[H]
    \centering
    \includegraphics[width=\linewidth]{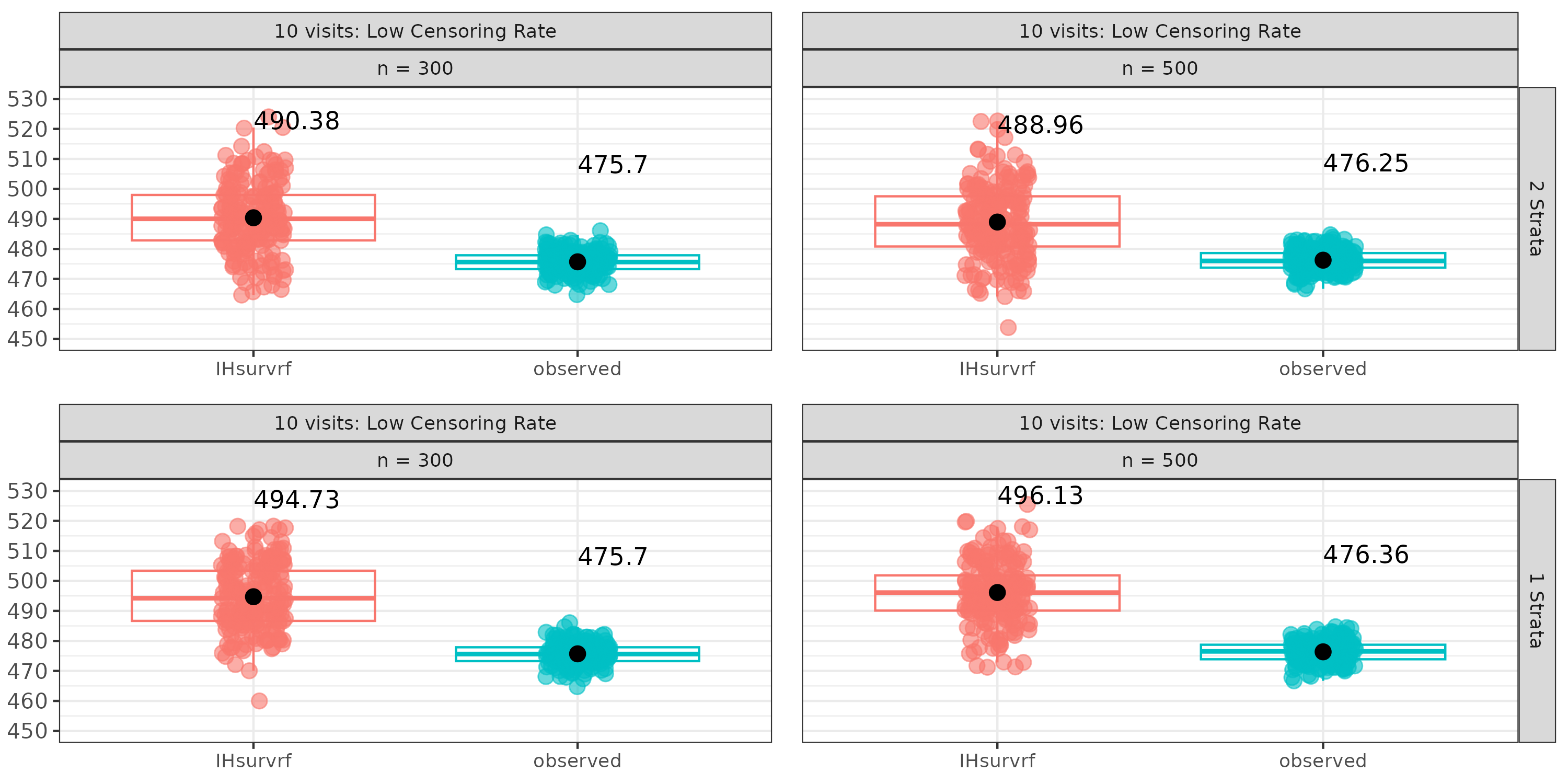}
    \caption{Boxplots for moderate censoring ($\sim$ 28\%) for 10 visits}
    \label{fig:enter-label}
\end{figure}

\begin{figure}[H]
    \centering
    \includegraphics[width=\linewidth]{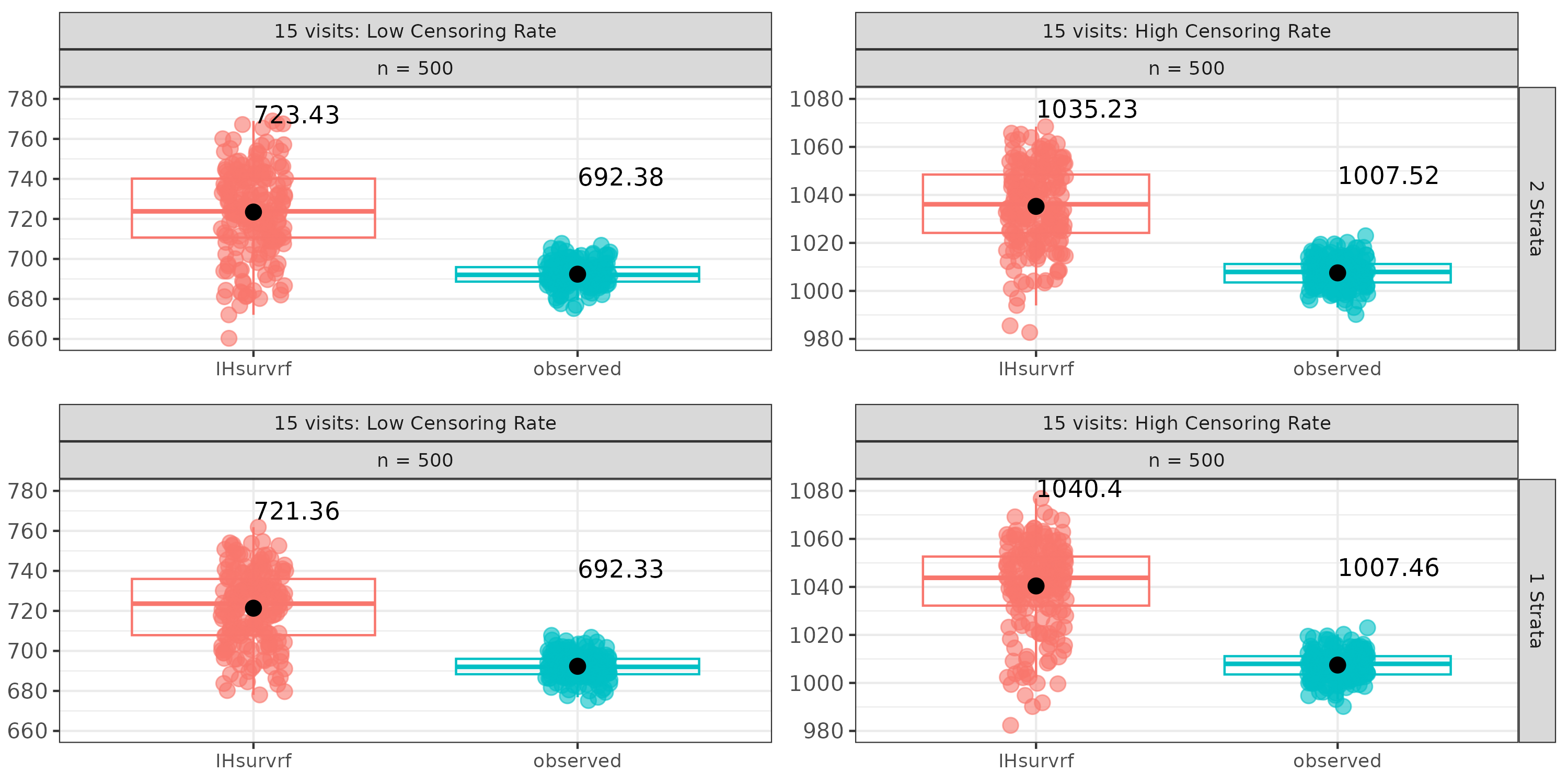}
    \caption{Boxplots for 15 visits, with both high ($\sim$ 50\%) and moderate ($\sim$ 30\%) censoring}
    \label{fig:enter-label}
\end{figure}

Each simulation replicate is a point, with a boxplot drawn with the median, Q1, and Q3. The mean is represented with a black point and the corresponding text label. The regime estimated using our new method is presented in red, and the observed regime is presented in blue. 

In each  setting, the proposed method outperforms the observed regime. However, a single strata forest results in marginally better performance than a $2$ strata forest in our simulated settings. This may be due to overfitting, since strata are defined by cumulative time rather than the number of visits experienced. 
This would especially be prevalent when patients have large differences in trajectories of their overall number of visits, as backwards recursion is performed visitwise.  


Computational times depend on the maximum number of visits, the sample size, and some tuning parameters such as the number of trees in a forest, minimum node size, and number of events in a node. 


\vspace{-2mm}

%% file: RDA.tex
{\setlength{\parindent}{15pt}

Crohn’s disease (CD) is a chronic, immune-mediated inflammatory disease of the gastrointestinal tract characterized by episodes of abdominal pain, bloody diarrhea, and malnutrition. CD impacts both pediatric and adult populations. CD is heterogeneous in both its presenting signs and symptoms as well as response to immune-suppressing therapies. Prospective studies of treatment effectiveness are limited in the pediatric population posing a significant challenge to treatment decision making for physicians and families. This includes postoperative pediatric patients in whom surgical resection of intestine is necessary given the severity of disease. Current postoperative pediatric practice guidelines are largely informed by expert consensus and limited retrospective data on account of a paucity of prospective interventional clinical trials.

We illustrate the use of this newly developed method using a cohort of pediatric patients with CD represented in the IQVIA Legacy PharMetric Administrative Claims Data. The IQVIA data include longitudinal, deidentified patient-level data comprised of inpatient, outpatient, and surgical encounters, as well as outpatient pharmacy claims from over 100 health plans and 50 million insurance plan enrollees. The IQVIA data have been used in prior CD epidemiologic research and are representative of commercially insured individuals in the United States \cite{barnes2021incidence, long2013increased, barnes2020decreasing}.

A clinical question of interest was: once a patient undergoes a first observed intestinal surgery (resection), when (and in what order?) should the physician start intervening to prescribe medication to maximize time-to-second resection, i.e. signifying good disease medical management and improved patient wellbeing? This falls under the umbrella of estimating an optimal dynamic treatment regime for such patients. We would typically begin patient follow-up from a patient’s first observed intestinal surgery and conclude follow-up at the time of primary outcome (repeat resection), or until the patient has been censored. However, since the event rate for a second surgery was only $10\%$ with a potentially large number of visits to reach the endpoint, we used a surrogate endpoint instead, time-to-second hospitalization. 

In pediatric patients with CD, maximizing time between hospitalizations, or even eliminating the need for hospitalization is an important objective in the management of disease. Treatment decision making is therefore driven by maximizing for this outcome however is informed by a paucity of real-worlddata. In this setting, remaining life is hospital-free survival. 

 A challenge in the application of our method to the intended indefinite horizon setting, is that standard IPCW cannot be applied accurately due to the large number of treatment visits that must match. This is an open research area to explore value function estimation for survival outcomes in indefinite horizon settings, with no current methods available, to our knowledge. For a proof-of-concept to illustrate that our method can accurately estimate a dynamic treatment regime, we apply the method to a setting with a maximum of $3$ visits, and apply standard IPCW techniques to estimate performance of our method against the observed policy.

 Baseline characteristics including age at first resection and sex were considered. Additionally, a maximum follow-up time of $1000$ days was considered, with administrative censoring occurring for patients with longer survival times. Similarly, a maximum of $3$ visits was considered for each patient. Other tailoring variables included the length of the previous visit, the number of previous visits, the counts of the number of times a patient was treated and untreated, and the cumulative time from baseline.


The final analysis was conducted using a cohort of $418$ patients who had at least 1 encounter with an International Classification of Diseases, 9th or 10th edition (ICD-9/10) code for CD, at least $1$ Current Procedural Terminology (CPT) code for intestinal resection, and had at least $1$ visit. $64$ patients experienced a second hospitalization after receiving a first resection within $3$ visits and the maximum study length, giving an event rate of $15.3\%$. $194$ ($47.41 \%$) of the patients were females, with the remaining $224$ ($53.59\%$) being males. We defined a binary treatment setting, where a patient either: a) received one common CD therapy including anti-tumor necrosis factor (aTNF) agents, aminosalicylates, immune modulators, steroids, anti-integrin agents, anti-interleukin agents, or antibiotics, or b) considered untreated.



Number of trees is a hyperparameter that can be tuned. We expect performance to increase with the number of trees until a certain threshold, with performance plateauing or declining after due to overfitting. For our dataset, this can be seen in Figure \ref{fig:val_trees}, where the blue line represents the performance of our method, and the grey line represents the value of the observed policy, which is what patients actually received. Performance of our method in the $2$ strata case peaks with $100$ trees.

\begin{figure}[H]
    \centering
\includegraphics[width=0.6\linewidth]{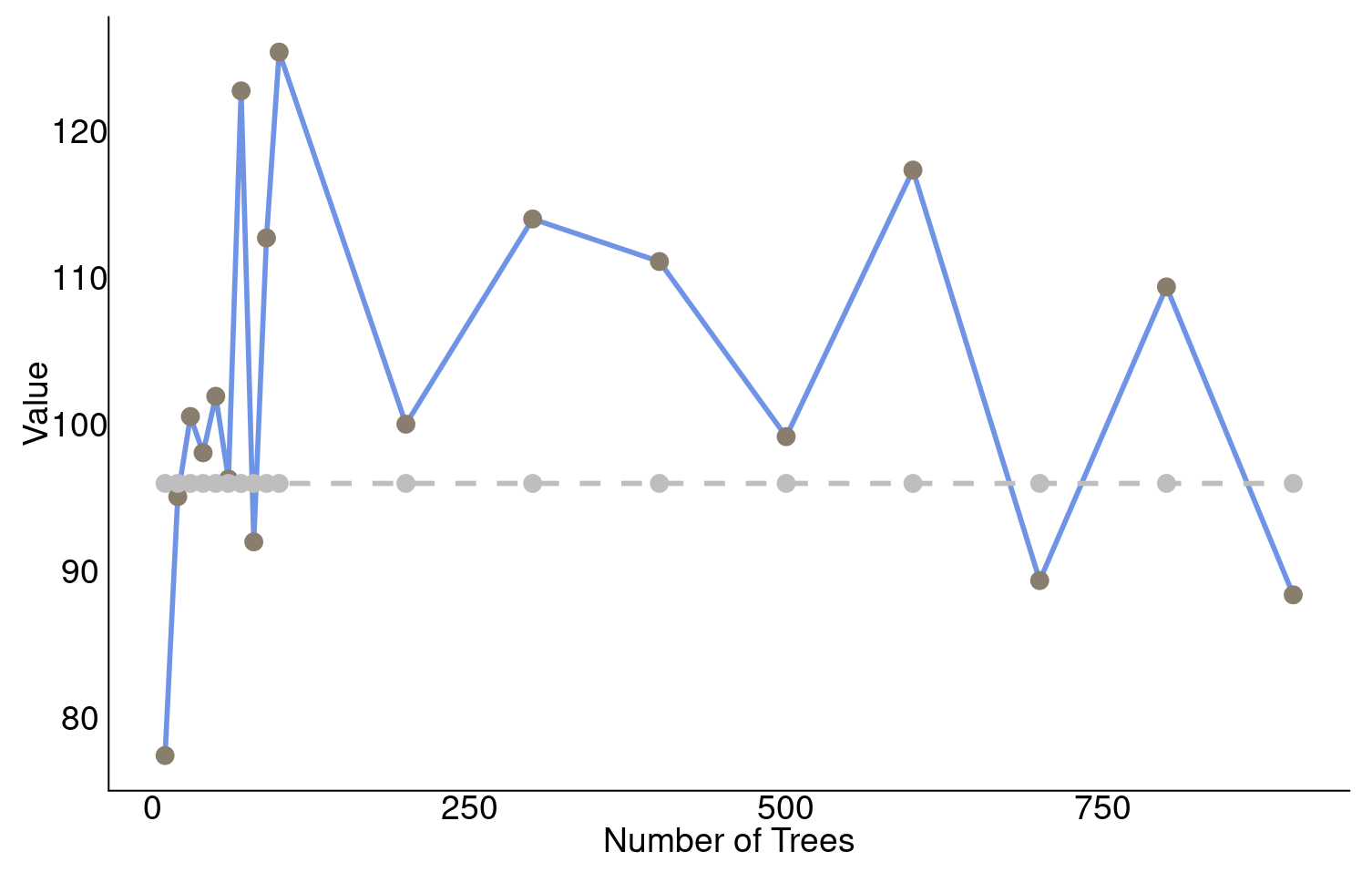}
    \caption{Performance of DTR estimator by number of trees (2 strata)}
    \label{fig:val_trees}
\end{figure}

\begin{figure}[H]
    \centering
\includegraphics[width=0.6\linewidth]{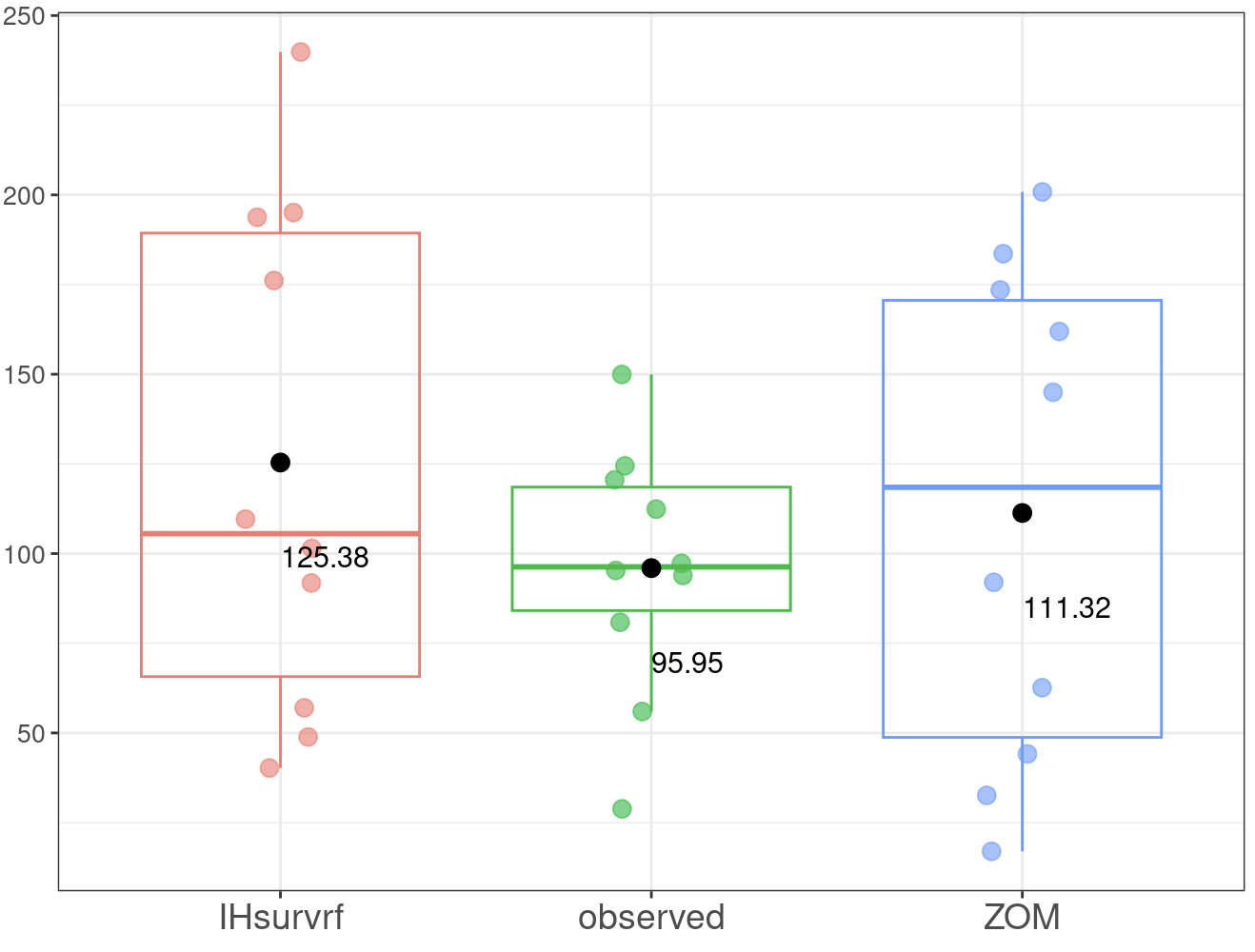}
        \caption{Performance using 100 trees}
    \label{fig:2strat_trees}
\end{figure}

We can plot the results of the analysis using $100$ trees in Figure \ref{fig:2strat_trees}. We compare our model with the estimated optimal zero-order model, where we don't consider patient heterogeneity and maximize population-wide average outcomes. We can see that the estimated regime outperforms the observed policy, and elongates time-to-second hospitalization by $\sim 30$ days. An additional limitation of our analysis is the presence of high censoring rates.

We illustrate the results of using the $2$ strata option, and examine performance against the number of trees used in the forest. Other hyperparameters include minimum node size, which was set to $5$, and the minimum number of events per node, which was set to $2$. The supplementary material contains additional details on how the analysis was performed, as well as the results of using a $1$ strata approach. 

\vspace{-3mm}



 

%% file: disc.tex
We propose a novel dynamic treatment regime estimator for indefinite horizon settings involving survival outcomes while allowing for a flexible number of visits, treatment arms, and dependent censoring. Under some regularity conditions for the forest, we proved uniform consistency of the value function of our estimator, which requires only looking at first visit information. We also proved uniform consistency of our pooled survival random forest as well as characterized the maximum number of iterations needed for the model refitting process. 

Due to computational intensity, we only explored up to 2 strata in the simulation settings. It would be of interest to see whether there are any performance benefits of further division into three or more forests, particularly in populations with a more pronounced difference in survival trajectories to prevent model overfitting. However, this may only be feasible if the number of events is not small, or is well-spread out across time, as each strata must have a minimum number of events to fit survival models.


Additionally, the algorithm requires extensive computational time due to the refitting of the forests, with additional run time needed for larger sample sizes and more visits. This may be a limiting factor for future analyses, so future work may consider methods of improving computational speed, perhaps by using a single random draw for survival probability augmentation from the previous visit, instead of appending the entire future survival information in the form of a probability.



A limitation of the application of our method discussed earlier is the need for development of novel value function estimators.
A well-known challenge with a large number of treatment visits is the barrier to directly using IPCW for value function estimation, as it is unlikely for all received treatments to match the optimal estimated treatment. An interesting approach may be the development of a proportion-based matching estimator for survival outcomes, which extends \cite{ye2023stage} to the survival setting. This allows for partial regime matching and would be an efficient use of the data, allowing for cross-validation approaches. Other approaches may include kernel-based estimators that directly weight the kaplan-meier curve with a propensity score based on whether-or-not a patient received the estimated optimal regime at each visit.

%% file: Supplement.tex













\begin{longtable}{|l|p{0.8\linewidth}|}
    \caption{Summary of notation} \label{tab:notation} \\
    \hline
    \multicolumn{2}{|c|}{\textbf{Basic notation}} \\
    \hline
    $k$ & Treatment visit. 1, 2, ..., $\tilde{K}$ \\
    $\tilde{K}$ & Constant. Largest number of visits across all patients. \\
    $\tau$ & Study length \\
    $\tau_l$ & Length of each strata \\
    $n_k$ & Number of training examples at visit $k$ \\
    $F$ & Cutoff time point for cumulative time for strata membership. \\
    $j$ & Index for number of refitting iterations. $1, 2, ..., J$\\
    $l$ & Index for number of strata. $1, 2, ..., W$\\
    $M_l$ & maximum number of visits in strata $l$.\\
    $J_l$ & maximum number of iterations in strata $l$.\\
    \hline
    \multicolumn{2}{|c|}{\textbf{Predictors}} \\
    \hline
    $A_k$ & Action at visit $k$ \\
    $Z_k$ & Covariate information at visit $k$ \\
    $H_k$ & History at visit $k$ \\
    $d_k$ & Dimension of history $H_k$ at visit $k$ \\
    $\pi_k(H_k)$ & Decision rule at visit $k$ given $H_k$ \\
    \hline
    \multicolumn{2}{|c|}{\textbf{visit-wise outcomes}} \\
    \hline
    $T_k$ & Failure time from visit $k$ \\
    $U_k$ & Time to next treatment from visit $k$ \\
    $C_k$ & Censoring time from visit $k$ \\
    $V_k$ & Uncensored length of visit $k$. $V_k = T_k \wedge U_k$ \\
    $X_k$ & Observed length of visit $k$. $X_k = V_k \wedge C_k$ \\
    $\delta_k$ & Censoring indicator of visit $k$. $\delta_k = I(V_k \leq C_k)$ \\
    $\gamma_k$ & Treatment indicator of visit $k$. $\gamma_k = I(T_k \leq U_k)$ \\
    $B_k$ & Baseline time at visit $k$. $B_k = \sum_{k'=1}^{k-1} X_{k'}$, $k > 1$, $B_1 = 0$ \\
    $L_k$ & Remaining life at visit $k$. $L_k = T - B_k$ \\
    \hline
    \multicolumn{2}{|c|}{\textbf{Overall outcomes}} \\
    \hline
    $T$ & Overall failure time. $T = \sum_{k=1}^{K} V_k$ \\
    $C$ & Overall censoring time \\
    $\delta$ & Overall censoring indicator. $\delta = 1(T \leq C)$ \\
    \hline
    \multicolumn{2}{|c|}{\textbf{Theoretical settings}} \\
    \hline
    $LS, LG$ & Lipschitz constants of Assumption 3 \\
    $n_{\text{min}}$ & Minimum terminal node size in Assumptions 1 \\
    $\beta$ & Rate of minimum terminal node size in Assumption 1 \\
    $\alpha$ & Regular split constant in Assumption 2 \\
    $\varphi$ & Stratified random split constant in Assumption 2 \\
    $\zeta$ & Covariate density bound in Assumptions 4 \\
    $c_1$ & At-risk probability bound at $\tau$ in Assumption 3 \\
    \hline
    \multicolumn{2}{|c|}{\textbf{Survival functions and values}} \\
    \hline
    $S$ & Overall or generic failure survival function \\
    $S_k$ & Survival function of remaining life at visit $k$ \\
    $G$ & Overall or generic censoring survival function \\
    $\phi_(S)$ & Mean truncated survival time. $\phi_(S) = E[S \wedge \tau]$ \\
    \hline
\end{longtable}

\setlength{\parindent}{0pt}

\section*{Proof of Lemma 1}

\vspace{2mm}

We want to show that for each strata we have a finite number of iterations. We can first show that the number of visits in the strata is finite by examining each strata's value of $\tau_l$, where for the first strata, $\tau_{l = 1}$ can be defined as the time from the strata splitpoint until the overall $\tau$, and the second, $\tau_{l = 2}$ (in a two-strata setting) is defined from baseline until the strata splitpoint. Since $\tau$ is finite, and $0 < \tau_{l = 1}, \tau_{l = 2} < \tau$, then $\tau_{l = 1}, \tau_{l = 2}$ are finite as well. Without loss of generality, this can be applied to any number of finite strata.

\vspace{2mm}

We define the maximum number of visits in each strata as $M_l$, with $l$ being the strata indicator. We know $M_l$ is finite, since the maximum possible number of visits can be defined as

\[
M_l = \frac{\tau_l}{min[\mathcal{X}_l]}.
\]

Since both the numerator and denominator are finite, as established above, $M_l$ must be finite, using assumption 4.

In the two strata scenario, we will first start with proving strata 1 has a finite number of iterations, then strata 2. The same concept can be applied to any finite number of strata, as long as there is enough data within each strata. 

\vspace{2mm}

\begin{center}
    \textit{Strata 1}
\end{center}

 The goal of the method is to, at each strata, optimize over the restricted mean survival curve. By definition of the restricted mean survival curve for all $i, j, k$ denoting patients, iterations, and visits respectively, using strata 1's maximum length, $\tau_1$ as an example,
 
 \vspace{2mm}

 \[
 0 < \int_0^{\tau_1} \hat{S}^*_{ikj} \{ \cdot | H_{ik} = H, A_{ikj} = \hat{\pi}_{ikj}(H_{ik}) \} dt \leq \tau_1.
 \]

Meaning, the maximum number of iterations required for the strata, $J_1$, is the total number of iterations required to optimize the survival information over the entire strata length, $\tau_1$. We argue that not only is $J_1$ finite, but that \underline{$J_1 = M_{l = 1}$}.\\

In the first model fitting iteration $j=1$, where we initialize the pooled random forest, we output optimized survival curves for each patient $i$ and for each visit $k$. We note that the visit times are the censoring times, which can differentiated by the value of the censoring indicator. The optimized survival curves can be denoted as 

\[
\hat{S}^*_{ikj=1} \{ \cdot | H_{ik} = H, A_{ikj=1} = \hat{\pi}_{ikj=1}(H_{ik}) \}.
\]

\vspace{5mm} 

However, such a quantity only maximizes survival over each visit individually, without considering any long-term treatment effects. Therefore, we move forward to the first re-fitting iteration of the forest. \\

The first refitting iteration, $j = 2$ uses the predicted survival curves, which use stochastic augmentation to link 2 visits together, if a patient experienced a next visit. Meaning, each visit's survival information now spans a longer time frame, of the current observed visit length $t$ for visit $k = K -1$ and the next visit's observed length $X_{k+1, i}$. Therefore, the optimized survival curves now span information from 2 visits at most. Again, we output optimized survival curves, this time with index $j = 2$. However, this optimization does not yet span over the entire $\tau_{l = 1}$. \\

Next, the third model fitting iteration, $j = 3$ again uses the stochastic augmentation aspect, where each patient's survival now spans the current visit's observed visit length, $t$ for visit $k = K -1$ and the next 2 visits $X_{k, i}, X_{k + 1, i}$, spanning information from 3 visits at most. \\

Using the same logic for the rest of the iterations, we can see that in order to cover the entire strata length, $\tau_{l = 1}$ which has a maximum of $M_1$ visits, we will require $M_1$ iterations at most. It should be noted, that no patient prediction may go beyond $\tau_{l = 1} $.

Thus, $J_1 = M_1$ in strata 1.

\begin{center}
    \textit{Strata 2}
\end{center}

Similarly for strata 2, the goal is to optimize over the restricted mean survival curve. We can split patients into 2 cases: (1) entire patient trajectory in strata 2, (2) some visits in strata 1. We prove in both cases, that the maximum number of iterations for strata 2 \underline{$J_2 = M_2$}. \\

For patients in the first case (1), the argument is simple, and matches that of the argument in strata 1, replacing $\tau_{l = 1}$ with $\tau_{l = 2}$, which is strata 2's maximum length, and can be defined by the length of time from baseline until the cutpoint dividing the strata. \\

For patients in the second case (2), the argument is similar. However, we note that the augmented survival curves will use the optimized next-visit survival probability, which may require using the information from strata 1. Thus, this no longer becomes a problem of optimizing the survival curve purely over $\tau_{l = 2}$, but over the entire $\tau$. This is because any survival information used from strata 1 inherently optimizes over the entire $\tau_{l = 1}$ space. \\

However, in a similar fashion, the first reffitting iteration $j = 2$ uses stochastic augmentation linking 2 visits together, where the later visit has optimized survival information over the entire $\tau_{l = 1}$. therefore, the current visit's survival information will span over the current visit length, $t$, which is part of $\tau_{l = 2}$'s length, and $\tau_{l = 1}$. \\

The second refitting iteration which is the third model fitting iteration $j = 3$ again uses stochastic augmentation, where the patient's survival spans the current observed visit length, $t$ for visit $k = K - 1$, the next visit, $X_{k, i}$, both of which are in $\tau_{l = 2}$'s length, and $\tau_{l = 1}$ using the optimized survival information from strata 1. \\

In order to then cover the entire length of $\tau$, which has a maximum of $M_1 + M_2$ visits, we will require $M_2$ iterations for strata 2, but overall, we will need $J = M_1 + M_2$. 

\begin{center}
    \textit{Strata W}
\end{center}

Without loss of generality, we can generalize the results to the $W$-th strata, where $W$ is finite. Using similar arguments, the goal of the $W$-th strata is to optimize the restricted mean survival curve by $\tau_W$ which is the maximum length of strata $W$. \\

The first refitting iteration $j = 2$ (after the pooling step) will use stochastic augmentation, linking 2 visits together using the final visit in strata $L$, and the first visit in strata $W - 1$, where the first visit in strata $W - 1$ has already optimized over the entire $\tau_{W-1}$. \\

In order to cover the entire length of $\tau$, which has a maximum of $M_1 + M_2 +  ... M_W$ visits, since each strata has a maximum number of visits, we will require \underline{$J = \sum_l M_l$} iterations for the entire algorithm, but $J_W = M_W$ iterations to optimize over strata $W$ alone. 

$\Box$ \\ 

\begin{figure}[H]
    \centering
    \includegraphics[width=\linewidth]{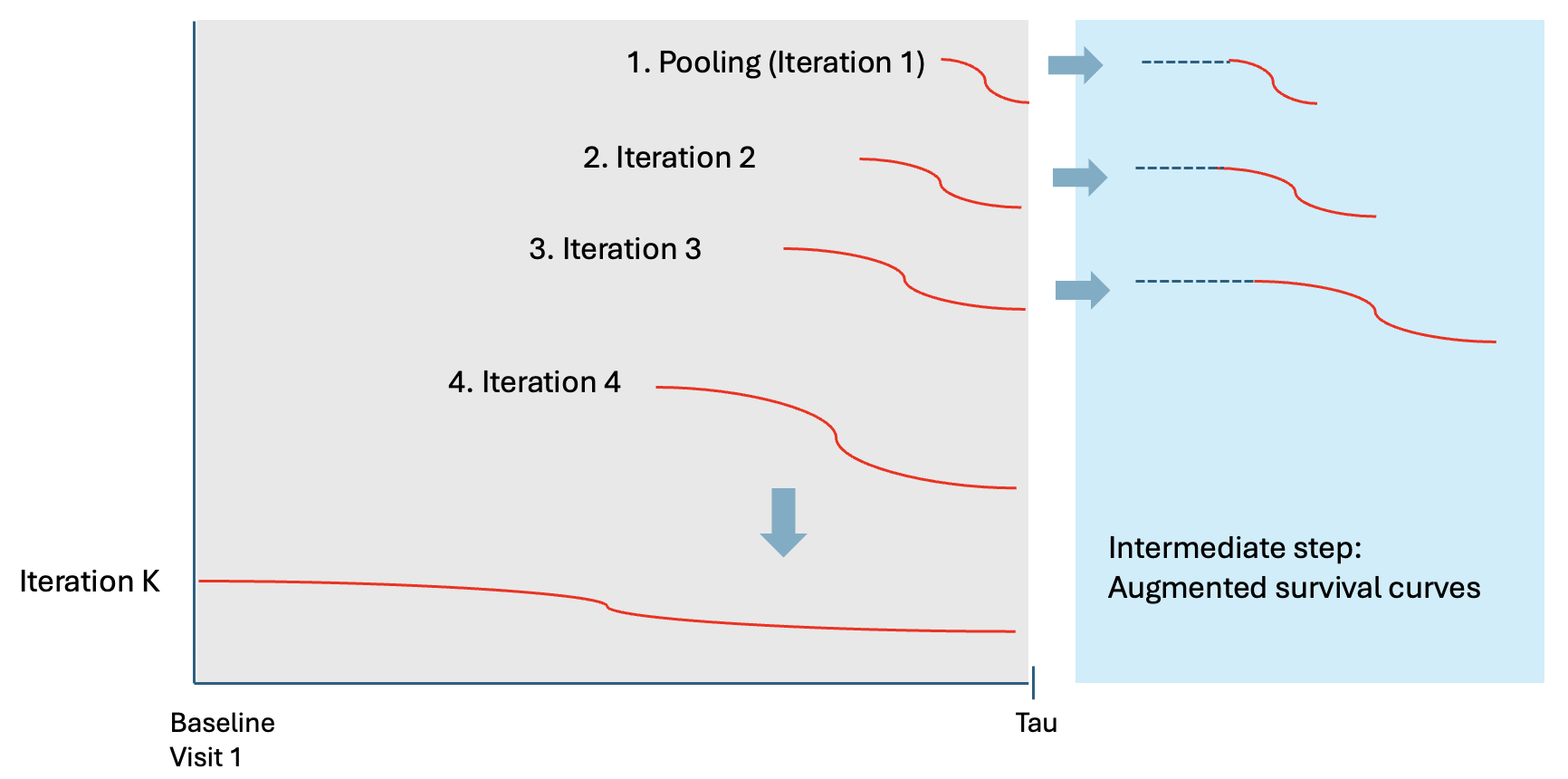}
    \caption{Illustration of Number of Required Iterations}
    \label{fig:proof_iter}
\end{figure}

As a visual example, please refer to Figure \ref{fig:proof_iter}, which illustrates the refitting process in a simple single strata example. In this case, take $K = \tilde{K}$ to be the maximum number of visits, and for each output survival curve to represent a single patient's optimized survival curve. The initial pooling only makes use of visit $\tilde{K}$'s information. At iteration $j = 2$ (after pooling), we then represent visit $\tilde{K}-1$'s optimized survival information, which accounts for visit $\tilde{K}$ as well. We will require $\tilde{K}$ iterations total to maximize the patient survival trajectory over the entire $\tau$ from visit $1$ to visit $\tilde{K}$.

\section*{Proof of Theorem 1}

\input{Supplement_Proofs/theorem1.tex}

\section*{Proof of Theorem 2}

{\setlength{\parindent}{15pt}

As Theorem 2 utilizes backwards recursion in optimization of the truncated mean survival criterion, \(\phi\), we first need to establish the validity of backwards recursion.

We begin by discussing the probabilistic augmentation step in detail. After the pooled forest has been trained in step 1, we input \(\bigg\{ (H_k, A_k = \hat{\pi}_k(H), \delta_k, V_k) \bigg\}_{i=1}^{M_l}\) for eligible \(k = K\) in strata $l$ through the forest of interest to obtain predicted optimized survival information in the form of \(\hat{S}_k^\ast(\cdot | H_k)\). This is the first step, which predicts optimized survival information for the final visit $k = K$. This optimized survival curve is then carried back to the previous visit $k = K - 1$ for eligible patients, who survived onto visit $K$; the $i$th patient with visit length $X_{K-1}$ is then given a probabilistic augmentation, where $\hat{S}^\ast_{k+1}(t - X_{k, i} | H_{k+1} = h_{k+1, i})$ represents the survival probability of remaining life at the last visit. For patients who have already experienced failure or censoring, no augmentation to the predicted optimal survival curve \(\hat{S}_k^\ast(\cdot | H_k)\) is required. The process is then repeated backwards visit-by-visit until $k = 1$ has been completed. The re-fitted forest is then fit on $\bigg\{(H_k, \hat{\pi}_k, \delta_k, \hat{S}^\ast_{k+1}(\cdot - X_{k} | H_{k+1}))\bigg\}^{W_l}_{i=1}$ for eligible $k = 1, 2...K-1$ in the strata and $\bigg\{(H_k, \hat{\pi}_k, \delta_k, \hat{S}^\ast_{k}(\cdot - X_{k} | H_{k}))\bigg\}^{W_l}_{i=1}$ for eligible $k = K$ in the strata, which relies on the use of the augmented versions of the optimized survival information.

Thus, the proof of theorem 2 relies on the legitimacy of Proposition 6, which is adapted from \cite{cho2023multi} to our setting which uses only next visit optimality, rather than all subsequent visit optimality. However, we will show in the proof that by the end of the model fitting process, the next visit optimality will inherently capture all future visit optimality. 

\vspace{5mm}

\setlength{\parindent}{0pt}

\input{Supplement_Proofs/backwards_recursion.tex}

\input{Supplement_Proofs/theorem2.tex}

\section*{Proof of Theorem 3}

\input{Supplement_Proofs/value_supplement.tex}

\section*{Additional Simulation Details}

The tables of coefficients used for the simulation scenarios are provided below:

\begin{table}[h!]
\footnotesize
\centering
\begin{tabular}{|c|c|c|}
\hline
\text{Beta} & Moderate Censoring & Higher Censoring  \\
\hline
$\beta_{T_k}$ & $(-5.5, -0.2, -0.5, -0.025, -0.02, 0.1, -0.08, 0.05)^T$  & $(-6, -0.2, -0.5, -0.025, -0.02, 0.1, -0.08, 0.05)^T$\\
$\beta_{U_k}$ & $(-1.5, -0.2, -0.5, -0.025, -0.02, 0.1, -0.08, 0.05)^T$ & $(-1.5, -0.2, -0.5, -0.025, -0.02, 0.1, -0.08, 0.05)^T$ \\
$\beta_{C_k}$ & $(-10.5, -0.2, -0.5, -0.025, -0.02, 0.1, -0.08, 0.05)^T$ & $(-6, -0.2, -0.5, -0.025, -0.02, 0.1, -0.08, 0.05)^T$ \\
\hline
\end{tabular}
\caption{Beta coefficients for $\tilde{K} = 10$ with N = 300 samples and $\tau = 1000$}
\end{table}

\begin{table}[h!]
\footnotesize
\centering
\begin{tabular}{|c|c|c|}
\hline
\text{Beta} & Moderate Censoring & Higher Censoring \\
\hline
$\beta_{T_k}$ & $(-5.5, -0.2, -0.5, -0.025, -0.02, 0.1, -0.08, 0.05)^T$ & $(-6, -0.2, -0.5, -0.025, -0.02, 0.1, -0.08, 0.05)^T$\\
$\beta_{U_k}$ & $(-1.5, -0.2, -0.5, -0.025, -0.02, 0.1, -0.08, 0.05)^T$ & $(-1.5, -0.2, -0.5, -0.025, -0.02, 0.1, -0.08, 0.05)^T$\\
$\beta_{C_k}$ & $(-12, -0.2, -0.5, -0.025, -0.02, 0.1, -0.08, 0.05)^T$ & $(-6, -0.2, -0.5, -0.025, -0.02, 0.1, -0.08, 0.05)^T$\\
\hline
\end{tabular}
\caption{Beta coefficients for $\tilde{K} = 10$ with N = 500 samples and $\tau = 1000$}
\end{table}

\begin{table}[h!]
\footnotesize
\centering
\begin{tabular}{|c|c|c|}
\hline
\text{Beta} & Moderate Censoring & Higher Censoring \\
\hline
$\beta_{T_k}$ & $(-5.5, -0.2, -0.5, -0.025, -0.02, 0.1, -0.08, 0.05)^T$ & $(-6.5, -0.2, -0.5, -0.025, -0.02, 0.1, -0.08, 0.05)^T$\\
$\beta_{U_k}$ & $(-1.5, -0.2, -0.5, -0.025, -0.02, 0.1, -0.08, 0.05)^T$ & $(-1.5, -0.2, -0.5, -0.025, -0.02, 0.1, -0.08, 0.05)^T$\\
$\beta_{C_k}$ & $(-11, -0.2, -0.5, -0.025, -0.02, 0.1, -0.08, 0.05)^T$ & $(-12, -0.2, -0.5, -0.025, -0.02, 0.1, -0.08, 0.05)^T$\\
\hline
\end{tabular}
\caption{Beta coefficients for $\tilde{K} = 15$ with N = 500 samples and $\tau = 1500$}
\end{table}

At the initial visit, each patient has two random state variables generated, $S_{1k}, S_{2k} \sim U(0, 1)$. Subsequent stages have state variables dependent on the value of the prior state, and are generated using the following equations; 

\vspace{-4mm}

\begin{align*}
S_{1,k+1} &= [\rho \cdot S_{1k} + \mathbf{1}_{ S_{1k} > 0.5} \cdot (\Delta_0 - A_k \cdot \Delta_1) + \mathbf{1}_{S_{1k} \leq 0.5} \cdot (\Delta_0^2 + A_k \cdot \Delta_1^2) + \frac{1}{2}g(1-\rho^2)^{\frac{1}{2}} \cdot U(0, 1)] - 0,  \\
S_{2,k+1} &= [\rho \cdot S_{2k} + \mathbf{1}_{ S_{2k} > 0.5} \cdot (\Delta_0 - A_k \cdot \Delta_1) + \mathbf{1}_{S_{2k} \leq 0.5} \cdot (\Delta_0^2 + A_k \cdot \Delta_1^2) + \frac{1}{2}g(1-\rho^2)^{\frac{1}{2}} \cdot U(0, 1)] - 0,
\end{align*}

where $\rho = 0.5$, $\Delta_0 = 0$, $\Delta_1 = 1$, and $g = 1$.

\section*{Crohn's Disease Data Analysis: Additional Details}

\input{Supplement_Proofs/RDA_details.tex}




%% file: Supplement_Proofs/theorem1.tex
In proving Theorem 1, it is enough to establish the consistency results for a single tree. As a forest is an average of multiple trees, where the randomization process of the trees is independent of the data generative law,the corresponding results for a forest follow. This is due to the supremum error of averages being less than the average of supremum errors for uniform consistency.\\

More specifically, with $\hat{S} = \frac{1}{n_{\text{tree}}} \sum_{w}^{n_{tree}} \hat{S}_{\{w\}}$, and

\[
\sup_{t,h} \bigg|\hat{S}(t \mid h) - S(t \mid h) \bigg| \leq \frac{1}{n_{\text{tree}}} \sum_{w}^{n_{tree}} \sup_{t,h} \bigg| \hat{S}_{\{w\}}(t \mid h) - S(t \mid h) \bigg|.
\]

\vspace{5mm}

Therefore, it is sufficient to prove Theorem 1 for a single tree $\hat{S}_{\{w\}}$, and for the sake of notational simplicity, we drop the tree index $w$ throughout the proof. 

\vspace{5mm}

We first introduce a modified Z-estimator lemma using theorem 2.10 of \cite{kosorok2008introduction}. Let $\Psi : \Theta \to L$ and $\Psi_n : \Theta \to L$ be maps between two normed spaces, where $\Theta$ is a parameter space, $L$ is some normed space, $\| \cdot \|_L$ denotes the uniform norm, $\Psi$ is a fixed map, and $\Psi_n$ is a data-dependent map. $\Theta_n$ is a  data-dependent survival function space, where survival functions are piecewise constant over a hyper-rectangle, represented by a rectangular kernel $k_h$, in the feature space so that $S(t \mid k_h) = S(t \mid h)$ for all $h \in \mathcal{H}$. The rectangular kernel $k_h$ is defined below.

\vspace{5mm}

\textbf{Lemma 2 (Consistency of Z-estimators).} Let $\Psi(S_0) = 0$ for some $S_0 \in \Theta$, and assume $\|\Psi(S_n)\|_L \to 0$ implies $\|S_n - S_0\|_\infty \to 0$ for any sequence $\{S_n\} \in \Theta_n \subset \Theta$ for $\Theta_n \to \Theta$ as $n \to \infty$. Then, if $\|\Psi_n(\hat{S}_n)\|_L \to 0$ in probability for some sequence of estimators $\hat{S}_n \in \Theta_n$ and $\sup_{S \in \Theta_n} \|\Psi_n(S) - \Psi(S)\|_L \to 0$ in probability, $\|\hat{S}_n - S_0\|_\infty \to 0$ in probability, as $n \to \infty$.

\vspace{5mm}

\textit{\underline{Proof of Lemma 2.}} Suppose a sequence $\hat{S}_n \in \Theta_n$ satisfies that $\|\Psi_n(S_n)\|_L \to 0$ and $\sup_{S \in \Theta_n} \|\Psi_n(S) - \Psi(S)\|_L \to 0$ both in probability as $n \to \infty$. Then we have

\[
\|\Psi(\hat{S}_n)\|_L = \|\Psi(\hat{S}_n) + \Psi(\hat{S}_n) - \Psi(\hat{S}_n)\|_L 
\leq \|\Psi(\hat{S}_n) - \Psi_n(\hat{S}_n)\|_L + \|\Psi_n(\hat{S}_n)\|_L
\]
\[
\leq \sup_{S \in \Theta_n} \|\Psi(S) - \Psi_n(S)\|_L + \|\Psi_n(\hat{S}_n)\|_L = \sup_{S \in \Theta_n} \|\Psi_n(S) - \Psi(S)\|_L + \|\Psi_n(\hat{S}_n)\|_L 
\]

which tends to $0$ in probability as $n \to \infty$ by the continuous mapping theorem. Thus, by the assumption, $\|\hat{S}_n - S_0\|_\infty \to 0$. $\Box$

\vspace{5mm}

For our setting, let $\Theta$ be the space of all covariate-conditional survival functions. Define a normed space $L = D_{[-1,1]} \{[0, \tau] \times \mathbb{R}^d\}$, where $D_AB$ is the space of all right-continuous left-limit functions with range $A$ and support $B$. We add $0$ to denote the true survival function, or $S_{k,0}$. For notational convenience we drop the current visit index $k$ when the context is clear, as $S \equiv S_k$ is an arbitrary candidate for the visit $k$ survival function, $S_0 \equiv S_{k,0}$ is the visit $k$ true functions. 

\vspace{5mm}

We define the following estimating equations, and the following propositions for proving consistency of the initial pooled forest for any strata $l \in L$, where $L$ is finite.

\begin{align}
\psi_{S,t} &= \delta I(X > t) + (1 - \delta) \left( \frac{S(t | H)}{S(X | H)} \land 1 \right) - S(t | H) \tag{1} \\
\Psi(S) &= \frac{P \psi_{S,t} \delta_h}{P \delta_h} \equiv P \cdot |h \, \psi_{S,t} \tag{2} \\
\Psi_n(S) &= \frac{\mathbb{P}_n \psi_{S,t} k_h}{\mathbb{P}_n k_h} \equiv \mathbb{P}_n \cdot |k_h \, \psi_{S,t} \tag{3};
\end{align}

\vspace{5mm}

where $P$ is the population average of the function values, $P f = \int f(h)dP(h)$, and $\mathbb{P}_n$ is the sample average of function values, i.e.,
\[
\mathbb{P}_n f = \int f(h) \, d\mathbb{P}_n(h) = \frac{1}{n} \sum_{i=1}^{n} f(H_i).
\]

$S_{0} \in \Theta$ is the true survival probability of the  remaining life, which resides in another survival space $\Theta$, $H \in \mathcal{H}$ is the historical information available at the current visit. Also, $\delta_h(h') = I(h' = h)$ is the unnormalized Dirac measure, and $k_h$ the unnormalized tree kernel. To be more specific, $k_h(h') = I(h' \in L(h))$, where $L(h)$ is the terminal node of the tree that contains the point $h$. We use the term 'unnormalized' to mean that they are not multiplied by the sample (or the population) size. 

By Lemma 2 and the following propositions, (1-4), we can show $\| \hat{S} - S_0 \|_\infty \to 0$ in probability as $n \to \infty$.

\vspace{5mm}

\textbf{Proposition 1.} \(\Psi(S_0) = 0\) for \(S_0 \in \Theta\).

\vspace{5mm}

\textbf{Proposition 2.} Assume Assumptions 1 to 4. As \(n \to \infty\), \(\|\Psi(\hat{S}_n)\|_L \to 0\) implies \(\|\hat{S}_n - S_0\|_\infty \to 0\).

\vspace{5mm}

\textbf{Proposition 3.} Assume Assumptions 1 to 4. Let \(\hat{S}_n\) be the kernel-conditional modified Kaplan-Meier estimator, which is the modified Kaplan-Meier in \cite{cho2023multi} applied to the data with weights indicated by the tree kernel, \(k_h\). As \(n \to \infty\), \(\|\Psi_n(\hat{S}_n)\|_L \to 0\) in probability.

\vspace{5mm}

\textbf{Proposition 4.} Assume Assumptions 1 to 4. As \(n \to \infty\), \(\sup_{S \in \Theta_n} \|\Psi_n(S) - \Psi(S)\|_L \to 0\) in probability.

\vspace{5mm}

The proofs for the above propositions are outlined below, and reference theorem 2 in \cite{zhou2024optimal} which utilize the same estimating equations to prove uniform consistency of a forest in a single visit setting. We use these to prove uniform consistency in a multi-visit, pooled random forest setting. Though there are slight differences in notation and setting, the proofs can be followed without loss of generality.

\vspace{5mm}

\textit{\underline{Proof of Proposition 1.}}  \(\Psi(S_0) = 0\) for \(S_0 \in \Theta\).

This proof can be referenced in \cite{zhou2024optimal}'s proof of theorem 2 for details. $\Box$

\vspace{5mm}

\textit{\underline{Proof of Proposition 2.}} As \(n \to \infty\), \(\|\Psi(\hat{S}_n)\|_L \to 0\) implies \(\|\hat{S}_n - S_0\|_\infty \to 0\).

By hypothesis, \(\|\Psi(\hat{S}_n)\|_L = \sup_{t \in [0, \tau_{1j=1}], h \in H} |\Psi(\hat{S}_n)| = \sup_{t \in [0, \tau_{1j=1}], h \in H} |P \cdot_{|h} \psi_{\hat{S}_n,t}| \to 0\). 

We show that \(P \cdot_{|h} \psi_{\hat{S}_n,t} \to 0\) uniformly over \(h \in H\) and \(t \in [0, \tau_{1j=1}]\) would imply

\[
\sup_{t, h} u_n(t|h) \to 0,
\]

where 

\[
u_n = \varepsilon_n (u|H, h) \left( 1 - G(t|H = h) \right) + \int_0^t \varepsilon(c|H = h) dG(c|H = h).
\]

Without loss of generality, we can follow the steps outlined in \cite{zhou2024optimal}'s proof of theorem 2. $\Box$

\vspace{5mm}

\textit{\underline{Proof of Proposition 3.}} As \(n \to \infty\), \(\|\Psi_n(\hat{S}_n)\|_L \to 0\) in probability.

We utilize proof by induction on \(t \in [0, \tau_{1j=1}]\), starting wih

\[
\sum_{i=1}^{n} \psi_{\hat{S}_{n}, t}(X_i) = \sum_{i=1}^{n} \left( I(X > 0) + (1 - \delta_i) I(X \leq 0) \frac{S(0|H)}{S(X|H)} - S(0 | H) \right)
\]

\[
= \sum_{i=1}^{n} \bigg\{ I(X > 0) - S(0 | H) \bigg\} = \sum_{i=1}^{n}\{1 - 1\} = 0.
\]

Then if \( \sum_{i=1}^{n} \psi_{\hat{S}_{n},t^-}(X_i) = 0 \), this implies \( \sum_{i=1}^{n} \psi_{\hat{S},t}(X_i) = 0 \) for all \( 0 < t \leq \tau_{1j=1} \). The desired result follows. We follow the approach outlined in \cite{zhou2024optimal}'s proof of theorem 2, by re-writing $\psi_{\hat{S}_{n},t}$ using counting process notation with jump sizes, using the same modified kaplan-meier estimator as in \cite{cho2023multi}. $\Box$

\vspace{5mm}

\textit{\underline{Proof of Proposition 4.}} As \(n \to \infty\), \(\sup_{S \in \Theta_n} \|\Psi_n(S) - \Psi(S)\|_L \to 0\) in probability.

The proof of proposition 4 requires proposition 5, which uses entropy calculations and empirical process theory to show that the tree (and therefore the forest) kernels are Donsker, as well as $\psi$.

\vspace{5mm}

\textbf{Proposition 5.} The collections of the unnormalized tree and forest kernel functions are Donsker, where the tree kernels $k^{tree}(\cdot)$ are axis-aligned random hyper-rectangles, and the forest kernels $k^{forest}(\cdot)$ are the mean of arbitrarily many tree kernels (n$_{tree}$).

\vspace{5mm}

\textit{\underline{Proof of Proposition 5.}} The proof of proposition 5 can be found in \cite{cho2023multi}.

For proposition 4, we first show that the class of functions \( \bigg\{\psi_{S,t} : S \in \Theta_n, t \in [0, \tau_{1j=1}]\bigg\} \) is Donsker, by similar arugments in \cite{cho2023multi}. As all terms in this class of functions belong to uniformly bounded Donsker classes through Lemma 9.10 of \cite{kosorok2008introduction}, the set of functions is Donsker.

Therefore, with proposition 5 and the class of products of bounded Donsker classes being Donsker, all of \( \bigg\{ \psi_{S,t} \delta_h : S \in \Theta_n, t \in [0, \tau_{1j=1}], h \in H \bigg\}\) are Donsker. Similarly, all of \(\bigg\{ \psi_{S,t} k_h : S \in \Theta_n, t \in [0, \tau_{1j=1}], h \in H \bigg\}\) are Donsker. By Assumption 1, \(\sup_h \mathbb{P}_n k_h \asymp n^{\beta - 1}\). 

We follow the proof outlined in \cite{zhou2024optimal}'s proof of theorem 2 without loss of generality, where $t \in [0, \tau_{1j=1} ]$ in our case.  $\Box$

\vspace{5mm}

%% file: Supplement_Proofs/backwards_recursion.tex
\textbf{Proposition 6 (Legitimacy of backward recursion)}. Let \( \{\pi\} \) be a non-arbitrary class of decision rules and let \( \pi^* = \{\pi_1^*, \pi_2^*, \dots, \pi_J^*\}^\top \) be a set of decision rules for each iteration $j = 1, ..., J$ which satisfy \( \phi \{S(\cdot \mid H_1, \pi^*)\} \geq \phi \{S(\cdot \mid H_1, \pi)\} \) for all \( \pi \in \{\pi\} \). Then, for any iteration \( j = 2, \dots, J \),

\[
\pi_j^*(H_j) = \arg \max_{a \in \mathcal{A}} \phi \bigg\{S_j(\cdot \mid H_j, \pi_{j-1}^*)\bigg\}
\]

Where $j$ represents each refitting iteration of the random forest relying on the previous iteration's estimates. In this case, we do not need to index each visit, $k$, separately for each iteration, as each forest is pooled across all visits $k$.

\underline{Proof of Proposition 6} 

We use a recursion strategy starting from iteration \( j = 1 \) and assume the following is true for some \( j = 1, 2, \dots, J \). Meaning, we are assuming optimatility of previous iterations, where each iteration covers all visits $k$ within the strata.

We will do this through inductive reasoning, where we first start with iteration $j = 1$.

\[
\pi_{j+1}^*(H_{j+1}) = \arg \max_{a \in \mathcal{A}} \phi \bigg\{ S_{j+1}(\cdot \mid H_{j+1}, \pi_{j}^* ) \bigg\}
\]

Then for any policy \( \pi_{j+1}^*(H_{j+1}) \) for the \( j \)-th iteration,

\[
\int_0^\tau S_{j+1}(t - X_j \mid H_{j+1}, A_{j+1} = \pi_{j+1}^*(H_{j+1})) dt
\]
\[
\geq \int_0^\tau S_{j+1}(t - X_{j} \mid H_{j+1}, A_{j+1} = \pi_{j+1}(H_{j+1})) dt.
\]

Thus for any $a \in \mathcal{A}$
\[
\phi \bigg\{ S_j(\cdot \mid H_j, a_j, \pi_{j+1}^*) \bigg\} 
\]
\[ = 
\int_0^\tau \int S_{j+1}\left( t - X_j \mid H_{j+1}, A_{j+1} = \pi_{j+1}^*(H_{j+1}) \right) 
dP(X_j, \delta_j, H_{j+1} \mid H_j, A_j = a_j) \, dt
\]
\[
= \int \int_0^\tau S_{j+1}\left( t - X_j \mid H_{j+1}, A_{j+1} = \pi^*_{j+1}(H_{j+1}) \right) dt \,
dP(X_j, \delta_j, H_{j+1} \mid H_j, A_j = a_j) 
\]
\[
\geq \int \int_0^\tau S_{j+1}\left( t - X_j \mid H_{j+1}, A_{j+1} = \pi_{j+1}(H_{j+1}) \right) dt \,
dP(X_j, \delta_j, H_{j+1} \mid H_j, A_j = a_j) 
\]
\[ = 
\phi \bigg\{ S_j(\cdot \mid H_j, a_j, \pi_{j+1}) \bigg\}.
\]

Meaning, after the $j = 1$ iteration, the optimal decision rule for $j = 2$ relies on optimizing the survival curves over two visit lengths. 

\vspace{5mm}

Next, for the $j = 2$ iteration, the optimal decision rule for $j = 3$ relies on optimizing the survival curves over three visit lengths, since we have optimized survival information from iteration $j = 2$, which already includes optimized information over iterations $j = 2$ and $j = 1$.

\vspace{5mm}

For the final iteration $j = J$, we will optimize the survival curves throughout the patient trajectory, as we will rely on using the previous iteration's ($j = J - 1$) optimized survival information, which inherently contains all the information from the previous iteration. More specifically, at the last iteration, all future visits until censoring are optimal.

\vspace{5mm}

Therefore, backwards recursion using augmented predicted optimal survival curves from the previous iteration in step 2 of the algorithm is valid. $\Box$

%% file: Supplement_Proofs/theorem2.tex
\vspace{5mm}

Similar to the approach for proving Theorem 1, it is enough to establish the consistency results for a single tree, as the forest is an average of multiple trees. The justification uses the same supremum error argument as theorem 1. Therefore, we will prove Theorem 2 for a single tree, $\hat{S}_{\{w\}}$, and drop the tree index $w$ for notational simplicity. 

Using Lemma 2 presented earlier in the proof of theorem 1 and the same notational setting, we define the following estimating equations:

\begin{align}
\psi_{S,t} &= \delta S_{k+1, 0}(t - x | H_{k + 1}) + (1 - \delta) \left( \frac{S(t | H)}{S(X | H)} \land 1 \right) - S(t | H) \tag{1} \\
\tilde{\psi}_{S,t, \tilde{S}_{k+1}} &= \delta \tilde{S}_{k+1, 0}(t - x | H_{k + 1}) + (1 - \delta) \left( \frac{S(t | H)}{S(X | H)} \land 1 \right) - S(t | H) \tag{2} \\
\Psi(S) &= \frac{P \psi_{S,t} \delta_h}{P \delta_h} \equiv P \cdot |h \, \psi_{S,t} \tag{3} \\
\Psi_n(S) &= \frac{\mathbb{P}_n \psi_{S,t} k_h}{\mathbb{P}_n k_h} \equiv \mathbb{P}_n \cdot |k_h \, \psi_{S,t} \tag{4};
\end{align}

we define the population average of the function values, $P$, and the sample average of function values $\mathbb{P}_n$ in the same manner as before.

Then, by Lemma 2, and the following propositions, (7-10), we can show $\| \hat{S} - S_0 \|_\infty \to 0$ in probability as $n \to \infty$.

\vspace{5mm}

\textbf{Proposition 7.} \(\Psi(S_0) = 0\) for \(S_0 \in \Theta\).

\vspace{5mm}

\textbf{Proposition 8.} Assume Assumptions 1 to 5. As \(n \to \infty\), \(\|\Psi(\hat{S}_n)\|_L \to 0\) implies \(\|\hat{S}_n - S_0\|_\infty \to 0\).

\vspace{5mm}

\textbf{Proposition 9.} Assume Assumptions 1 to 5. Let \(\hat{S}_n\) be the kernel-conditional modified Kaplan-Meier estimator, which is the modified Kaplan-Meier in \cite{cho2023multi} applied to the data with weights indicated by the tree kernel, \(k_h\). As \(n \to \infty\), \(\|\Psi_n(\hat{S}_n)\|_L \to 0\) in probability.

\vspace{5mm}

\textbf{Proposition 10.} Assume Assumptions 1 to 5. As \(n \to \infty\), \(\sup_{S \in \Theta_n} \|\Psi_n(S) - \Psi(S)\|_L \to 0\) in probability.

\vspace{5mm}

The proofs for the above propositions are outlined below, and reference \cite{cho2023multi} which utilize the same estimating equations to prove uniform consistency of a forest using augmented survival information from the next visit. We expand upon the proofs provided in the original paper for clarity.

\vspace{5mm}

\textit{\underline{Proof of Proposition 7.}}  \(\Psi(S_0) = 0\) for \(S_0 \in \Theta\).

\[
\Psi(S_0) = \frac{P \psi_{S_0,t} \delta_h}{P \delta_h} = P( \delta_h \bigg[\delta S_{k+1, 0}(t - x | H_{k+1}) + (1- \delta)\bigg\{ \frac{S_0(t | H)}{S_0(X | H)} \land 1 \bigg\} - S_0(t | H)\bigg]) \]
\[
= 
E_{\psi_{S_0, t}\delta_h}( \delta_h \bigg[\delta S_{k+1, 0}(t - x | H_{k+1}) + (1- \delta)\bigg\{ \frac{S_0(t | H)}{S_0(X | H)} \land 1 \bigg\} - S_0(t | H) \bigg])
\]
\[
= 
E_{\delta_h} \bigg\{ E_{\psi_{S_0, t}\delta_h}( \delta_h \bigg[\delta S_{k+1, 0}(t - x | H_{k+1}) + (1- \delta)I(X > t) + (1 - \delta)I(X \leq t)\frac{S_0(t | H)}{S_0(X | H)} - S_0(t | H) \bigg] | \delta_h) \bigg\}
\]
\begin{align*}
= P(\delta_h = 1) \bigg\{ 
& E_{\psi_{S_0, t} \delta_h} \bigg[\delta S_{k+1, 0}(t - x | H_{k+1})\bigg] \tag{A} \\
& + E_{\psi_{S_0, t} \delta_h} \bigg[ (1 - \delta) I(X > t) \bigg] \tag{B} \\
& + E_{\psi_{S_0, t} \delta_h} \bigg[ (1 - \delta) I(X \leq t) \frac{S_0(t | H)}{S_0(X | H)} \bigg] \tag{C} \\
& - E_{\psi_{S_0, t} \delta_h} \bigg[ S_0(t | H) \bigg] \tag{D} 
\bigg\}.
\end{align*}

Going term-by-term,

\[ (A)= E_{\psi_{S_0, t} \delta_h} \bigg[\delta S_{k+1, 0}(t - x | H_{k+1})\bigg] =
 E_{\psi_{S_0, t} \delta_h}\bigg[I( C > t) S_{k+1} (t - x | H_{k + 1})\bigg].
\]

We note that $S_{k+1, 0}(t - x | H_{k+1})$ can be written as a piecewise function, taking value of $1$ from $t - x$ to $t$, and taking on survival function $S_0(t | H)$ above $t$. Therefore, we can rewrite the expected value substituting the piecewise survival function, where we know the censoring time is greater than $t$ since we have survived until the next visit as

\[
 E_{\psi_{S_0, t} \delta_h} \bigg[I(C > t) \{ I(t - X \leq T \leq t) + S_0(t | H) I(T > t) \} \bigg] 
\]
\[=
 \underbrace{E_{\psi_{S_0, t} \delta_h}\bigg[I(C > t)I(t - X \leq T \leq t)\bigg]}_{1} +  \underbrace{E_{\psi_{S_0, t} \delta_h}\bigg[I(C > t) S_0(t | H) I(T > t)\bigg]}_{2}.\]

The first term will be $0$, as the probability of having an event between $t 
- X$ and $t$ is $0$, as we know that no event has occurred in that time period due to the patient surviving until the next visit. \\ 

For term 2, $E_{\psi_{S_0, t} \delta_h}\bigg[I(C > t) S_0(t | H) I(T > t)\bigg] = S_0(t | H) E_{\psi_{S_0, t} \delta_h}\bigg[I(C > t) I(T > t)\bigg]$ 

\[
= S_0(t | H) \int_t^\infty \int_t^\infty f_{T, C}(u, c) dc du = S_0(t | H) \int_t^\infty \int_t^\infty f_{T}(u) f_C(c) dc du
\]
\[
= S_0(t | H)\{1 - G(t)\} \int_t^\infty f_T(u)du
\]
\[
= S_0(t | H)\{1 - G(t)\}\bigg[ F_t(\infty) - F_t(t) \bigg] = S_0(t | H)\{1 - G(t)\} \bigg[ 1 - P(T \leq t) \bigg] = S_0(t | H)\{1 - G(t).\}
\]

The last equality follows since $P(T \leq t) = 0$, since in this situation there are no events before time $t$, as a patient has survived until the next visit. \\


\[(B)=
E_{\psi_{S_0, t} \delta_h} \bigg[ (1 - \delta) I(X > t) \bigg] = 0,
\]

so if $\delta = 1$, then it follows that this is 0. If $\delta = 0$, then there is censoring, so $C \leq T$ and $X = C$ due to censoring. However, $T > t$ since we know the patient has survived until time $t$, this is not possible. Meaning, the overall expected value will be $0$.






\[(C)= 
E_{\psi_{S_0, t} \delta_h} \left[ (1 - \delta) I(X \leq t) \frac{S_0(t | H)}{S_0(X | H)} \right] 
\]
\[=
S_0(t|H) \int_c^\infty \frac{1}{S_0(C| H)} \int_0^{t_0} f_C(c)f_T(u)du
\]
\[ = S_0(t|H)F_C(c) |_{c = 0}^{c = t} =
S_0(t|H)\bigg[ F_C(t) - F_C(0)\bigg] = S_0(t|H)G(t),
\]

where $G(t)$ denotes the censoring distribution, i.e., $G(t) = P(C \leq t)$. Finally, 

\[(D)=
- E_{\psi_{S_0, t} \delta_h} \bigg[ S_0(t | H)\bigg] = -S_0(t | H).
\]

Putting everything together,

\vspace{-2mm}

\[
P(\delta_h = 1) \bigg\{ 
(A) +  (B) + (C) + (D)
\bigg\} =
P(\delta_h = 1) \bigg\{ 
S_0(t | H)\{1 - G(t)\} + S_0(t|H)G(t) - S_0(t | H)
\bigg\} 
\]
\[
 = 0. 
\]

$\Box$

\vspace{5mm}

\textit{\underline{Proof of Proposition 8.}} As \(n \to \infty\), \(\|\Psi(\hat{S}_n)\|_L \to 0\) implies \(\|\hat{S}_n - S_0\|_\infty \to 0\).

By hypothesis, \(\|\Psi(\hat{S}_n)\|_L = \sup_{t \in [0, \tau_{lj}], h \in \mathcal{H}} |\Psi(\hat{S}_n)| = \sup_{t \in [0, \tau_{lj}], h \in \mathcal{H}} |P \cdot_{|h} \psi_{\hat{S}_n,t}| \to 0\). 

We show that \(P \cdot_{|h} \psi_{\hat{S}_n,t} \to 0\) uniformly over \(h \in \mathcal{H}\) and \(t \in [0, \tau_{lj}]\) would imply

\[
\sup_{t, h} u_n(t|h) \to 0,
\]

where 

\[
u_n = \varepsilon_n (t|H = h) \left( 1 - G(t|H = h) \right) + \int_0^t \varepsilon(c|H = h) dG(c|H = h).
\]

\[
P \cdot_{|h} \psi_{\hat{S}_n,t} = P_{\cdot | h} \bigg[ \delta S_{k+1, 0}(t - x | H_{k+1}) + (1 - \delta) \bigg\{\frac{\hat{S}_n(t|H)}{\hat{S}_n(X | H)} \land 1 \bigg\}  - \hat{S}_n(t | H)\bigg]
\]
\begin{align*}
= & \, P_{\cdot | h} \bigg\{ \delta S_{k+1, 0}(t - x | H_{k+1}) \tag{A} \\
& \quad + (1 - \delta) I(X > t) \tag{B} \\
& \quad + (1 - \delta) I(X \leq t)\frac{\hat{S}_n(t|H)}{\hat{S}_n(X | H)} \tag{C} \\
& \quad - \hat{S}_n(t | H) \tag{D} \bigg\}.
\end{align*}

Now, we examine each term individually, starting with 

\[ (A)=
P_{\cdot | h}\bigg\{ \delta S_{k+1, 0}(t - x | H_{k+1}) \bigg\} = S_0(t | H = h) \{1 - G(t | H = h) \}.
\]

This proof is similar to the one from the proof of proposition 7 for term (A). We can see that

\[ (B)=
P_{\cdot | h}\bigg\{ (1 - \delta) I(X > t) \bigg\} = 0,
\]

because if $\delta = 1$, then this is 0. If $\delta = 0$ then $C \leq T$ and $X = C$ due to censoring, then $I(C > t) = 0$ since we know the patient has survived until time $t$, so $T > t$. \\

\[
(C)= P_{\cdot | h} \bigg\{ (1 - \delta) I(X \leq t)\frac{\hat{S}_n(t|H)}{\hat{S}_n(X | H)} \bigg\} = \int_0^\infty (1 - \delta) I(X \leq t)\frac{\hat{S}_n(t|H = h)}{\hat{S}_n(X | H = h)} dP_{\cdot | H = h}
\]
\[
= E\bigg[(1 - \delta) I(X \leq t)\frac{\hat{S}_n(t|H = h)}{\hat{S}_n(X | H = h)}\bigg] = \hat{S}_n(t | H = h) E\bigg[\frac{I(\delta = 0) I(X \leq t)}{\hat{S}_n(c|H = h)}\bigg]
\]
\[
= \int_0^t \frac{f_C(c) S_0(c | H = h)}{\hat{S}_n(c|H = h)}dc \hat{S}_n(t|H = h). 
\]
Finally, 

\[
(D)= P_{\cdot | h}\bigg\{ -\hat{S}_n(t | H) \bigg\} = - \int_0^\infty \hat{S}_n(t | H) dP_{\cdot | H = h} = - \hat{S}_n(t | H = h).
\]

Putting the terms together, we get 
\[
\{1 - G(t | H = h) \}S_0(t|H = h) + \int_0^t \frac{f_C(c) S_0(c | H = h)}{\hat{S}_n(c|H = h)}dc \hat{S}_n(t|H = h) - \hat{S}_n(t | H = h).
\]

We let $\varepsilon_n(t | H = h) = \frac{S_0(t | H = h)}{\hat{S}_n(t | H = h)} - 1$, and substitute this into the expression above.

\begin{align*}
&= \frac{\hat{S}_n(t | H = h)}{\hat{S}_n(t | H = h)} S_0(t | H = h) - \hat{S}_n(t | H = h) + \int_0^t \frac{f_C(c) S_0(c | H = h)}{\hat{S}_n(c | H = h)} dc \, \hat{S}_n(t | H = h) \\
&\quad - \frac{\hat{S}_n(t | H = h)}{\hat{S}_n(t | H = h)} G(t | H = h) S_0(t | H = h) + \hat{S}_n(t | H = h) G(t | H = h) - \hat{S}_n(t | H = h) G(t | H = h)
\end{align*}

\[
= \hat{S}_n(t | H = h) \bigg \{
\frac{S_0(t | H = h)}{\hat{S}_n(t | H = h)} - 1  + \int_0^t \frac{f_C(c) S_0(c | H = h)}{\hat{S}_n(c | H = h)} dc - G(t | H = h) \bigg[ \frac{S_0(t | H = h)}{\hat{S}_n(t | H = h)} - 1 \bigg] - G(t | H = h)
\bigg\}.
\]

Next, we note that $f_C(c) dc = dG(c | H = h)$ so,
\[
\int_0^t \frac{S_0(c | H = h)}{\hat{S}_n(c | H = h)} dG(c | H = h) - G(t | H = h) = \int_0^t \frac{S_0(c | H = h)}{\hat{S}_n(c | H = h)} dG(c | H = h) - \int_0^t dG(c | H = h)
\]
\[
 = \int_0^t \varepsilon_n(c | H = h) dG(c | H = h).
\]

Now, combining the terms together once again, we have
\[
= \hat{S}_n(t | H = h) \bigg \{
\frac{S_0(t | H = h)}{\hat{S}_n(t | H = h)} - 1  - G(t | H = h) \bigg[ \frac{S_0(t | H = h)}{\hat{S}_n(t | H = h)} - 1 \bigg] + \int_0^t \varepsilon_n(c | H = h) dG(c | H = h)
\bigg\}
\]
\[
= \hat{S}_n(t | H = h) \bigg \{
\int_0^t \varepsilon_n(c | H = h) dG(c | H = h) + \varepsilon_n(t | H = h) \bigg[1 - G(t | H = h) \bigg]
\bigg\}.
\]

We define $U_n(t|H = h) = \int_0^t \varepsilon_n(c | H = h) dG(c | H = h) + \varepsilon_n(t | H = h)\bigg[1 - G(t | H = h) \bigg]$.

Taking the derivative of both sides using the fundamental theorem of calculus, we get 
\[
dU_n(t|H = h) = \frac{d}{dt} \bigg\{
\int_0^t \varepsilon_n(c | H = h) dG(c | H = h) + \varepsilon_n(t | H = h)\bigg[1 - G(t | H = h) \bigg]
\bigg\}
\]
\[
= \varepsilon_n(t | H = h) dG(t | H = h) + \frac{d}{dt} \bigg\{ \varepsilon_n(t | H = h) \bigg[1 - G(t | H = h) \bigg] \bigg\}
\]
\[
= \frac{d}{dt} \varepsilon_n(t | H = h) \bigg[1 - G(t | H = h) \bigg].
\]

Now by rearranging the equation,
\[
\frac{d}{dt} \varepsilon_n(t | H = h) = \frac{d U_n(t | H = h)}{1 - G(t | H = h)}.
\]

Integrating both sides by t, we get \[
\epsilon_n(t | H = h) = \int_0^t \frac{d U_n(t | H = h)}{1 - G(t | H = h)}.
\]

To solve this, we can integrate by parts, setting $u = \frac{1}{1 - G(t | H = h)}$, $dv = \frac{d}{dt} U_n(t | H = h)$, resulting in

\[
\epsilon_n(t | H = h) = \frac{U_n(t | H = h)}{1 - G(t | H = h)} - \int_0^t \frac{U_n(t | H = h) dG(t | H = h)}{\{ 1 - G(t | H = h) \}^2 }.
\]

We know that $G(t | H = h) \leq G(\tau | H = h)$ since $G(t)$ is the censoring distribution, and $0 \leq P(C \leq t) \leq P(C \leq \tau)$ since $0 < t \leq \tau$. Therefore, $1 - G(t | H = h) \geq 1 - G(\tau | H = h) > 0$.

Thus, $U_n(t | H = h) \rightarrow 0$ implies $\varepsilon_n(t | H = h) \rightarrow 0$. By hypothesis, we have that $U_n(t | H = h) \rightarrow 0$, therefore, $\varepsilon_n(t | H = h) \rightarrow 0$, and $\sup U_n(t | H = h) \rightarrow 0$. Thus, \(\|\hat{S}_n - S_0\|_\infty \to 0\)

$\Box$

\vspace{5mm}

\textit{\underline{Proof of Proposition 9.}} As \(n \to \infty\), \(\|\Psi_n(\hat{S}_n)\|_L \to 0\) in probability.

We utilize proof by induction on \(t \in [0, \tau_{lj}]\), starting with 

\[
\sum_{i=1}^{n} \psi_{\hat{S}_{n}, t = 0}(X_i) = \sum_{i=1}^{n} \left( \delta_i S_{k + 1, 0}(- X_i | H_{k + 1})  + (1 - \delta_i) I(X_i > 0) + \underbrace{(1 - \delta_i)I(X_i \leq 0)\frac{\hat{S}(0|H)}{\hat{S}(X_i|H)}}_{\text{A}} - \underbrace{\hat{S}(0 | H)}_{\text{B}} \right).
\]

Going term-by-term,
\[
(A)= \text{We note that in survival analysis, $X$ cannot be $0$. Therefore, $I(X_i \leq 0) = 0$}.
\]

\[
(B)= \quad \hat{S}(0 | H) = 1 \quad \text{since at time } 0, \text{ the survival probability is } 1.
\]

Now looking at the remaining two terms, there are $2$ possible cases depending on the value of $\delta$.

\[
\underline{\text{Case 1:}} \ \delta = 0; \quad \text{the expression becomes:} \quad \sum_{i = 1}^n \{ I(X_i > 0) - 1 \} = \sum_{i = 1}^n \{ 1 - 1 \} = 0.
\]

\[
\underline{\text{Case 2:}} \ \delta = 1; \quad
\text{the expression becomes} \quad \sum_{i = 1}^n \{S_{k+ 1, 0}(- X_i | H_{k + 1}) - 1 \} = 0.
\]

Then, if $\sum_{i = 1}^n \psi_{\hat{S}_{n}, t-}(X_i) = 0$ implies $\sum_{i = 1}^n \psi_{\hat{S}_{n}, t}(X_i) = 0$  for all $0 < t \leq \tau_{lj}$, the desired result follows. Now, we examine $\sum_{i=1}^{n} \psi_{\hat{S}_{n}, t} $

\[
= \sum_{i=1}^{n} \left( \delta_i S_{k + 1, 0}(t - X_i | H_{k + 1})  + (1 - \delta_i) I(X_i > t) + (1 - \delta_i)I(X_i \leq t)\frac{\hat{S}_n(t|H)}{\hat{S}_n(X_i|H)} - \hat{S}_n(t | H) \right).
\]

We know for a function, $f(t) = f(t-) + df(t)$, where $df(t)$ represents the jump size at time $t$. Therefore, we re-write the expression using jumps.

\begin{align*}
& \sum_{i=1}^{n} \psi_{\hat{S}_{n}, t} = \sum_{i = 1}^n \delta_i S_{k + 1, 0}(t^- - X_i | H_{K + 1}) + \sum_{i = 1}^n \delta_i dS_{k + 1, 0}(t - X_i | H_{K + 1}) \\
& + \sum_{i = 1}^n (1 - \delta_i) I(X_i > t^-) + \sum_{i = 1}^n (1 - \delta_i) dI(X_i > t) \\
& + \sum_{i = 1}^n \left\{ \frac{(1 - \delta_i) I(X_i \leq t^-)}{\hat{S}_n(X_i)} - 1 \right\} \hat{S}_n(t) + \sum_{i = 1}^n \left\{ \frac{(1 - \delta_i) dI(X_i \leq t^-)}{\hat{S}_n(X_i)} \right\} \hat{S}_n(t).
\end{align*}

From the definition of the modified kaplan-meier, $\hat{S}_k(t) = \Pi_{s \geq 0}^t \bigg\{ 1 + \frac{\sum_i \delta_{k, i} dS_{k, i}(s)}{\sum_i S_{k, i}(s^-)} \bigg\}$, where $S_{k, i}(t) = I(t > X_{k, i})$ and $\delta = 1$ indicates no censoring.

This expression becomes 
\[
\hat{S}_k(t) = \Pi_{s \geq 0}^t \bigg\{ 1 + \frac{\sum_i \delta_{k, i} dI(X_{k, i} > t)}{\sum_i I(X_{k, i} > t^-)} \bigg\}.
\]

Therefore by definition,

\[
\hat{S}_k(t) = \hat{S}_k(t^-) \bigg\{ 1 + \frac{\sum_i \delta_{k, i} dI(X_{k, i} > t)}{\sum_i I(X_{k, i} > t^-)} \bigg\},
\]

so

\[
\hat{S}_n(t) = \hat{S}_n(t^-) + \hat{S}_n(t^-) \bigg\{ \frac{\sum_i \delta_{k, i} dI(X_{k+1, i} > t)}{\sum_i I(X_{k+1, i} > t^-)} \bigg\}.
\]

We can then rewrite the expression as

\begin{align*}
& \sum_{i=1}^{n} \psi_{\hat{S}_{n}, t} = \sum_{i = 1}^n \delta_i S_{k + 1, 0}(t^- - X_i | H_{k + 1}) + \sum_{i = 1}^n \delta_i dS_{k + 1, 0}(t - X_i | H_{k + 1}) \\
& + \sum_{i = 1}^n (1 - \delta_i) I(X_i > t^-) + \sum_{i = 1}^n (1 - \delta_i) dI(X_i > t) \\
& + \sum_{i = 1}^n \left\{ \frac{(1 - \delta_i) I(X_i \leq t^-)}{\hat{S}_n(X_i)} - 1 \right\} \hat{S}_n(t^-) + \sum_{i = 1}^n \left\{ \frac{(1 - \delta_i) I(X_i \leq t^-)}{\hat{S}_n(X_i)} - 1 \right\} \hat{S}_n(t^-) \frac{\delta_i dI(X_{k + 1, i} > t)}{I(X_{k + 1, i} > t^-)} \\
& + 
\sum_{i = 1}^n \left\{ \frac{(1 - \delta_i) dI(X_i \leq t)}{\hat{S}_n(X_i)}\right\} \hat{S}_n(t).
\end{align*}

By hypothesis, $\sum_{i=1}^{n} \psi_{\hat{S}_{n}, t^-} = 0 \\ 
= \sum_{i = 1}^n \delta_i S_{k + 1, 0}(t^- - X_i | H_{K + 1}) + \sum_{i = 1}^n (1 - \delta_i) I(X_i > t^-) + \sum_{i = 1}^n \left\{ \frac{(1 - \delta_i) I(X_i \leq t^-)}{\hat{S}_n(X_i)} -1 \right\} \hat{S}_n(t^-)$.

Substituting this result, we get

\begin{align*}
& \sum_{i=1}^{n} \psi_{\hat{S}_{n}, t} = 
\sum_{i = 1}^n \delta_i dS_{k + 1, 0}(t - X_i | H_{K + 1}) + \sum_{i = 1}^n (1 - \delta_i) dI(X_i > t) \\
& + \sum_{i = 1}^n \left\{ \frac{(1 - \delta_i) I(X_i \leq t^-)}{\hat{S}_n(X_i)} - 1 \right\} \hat{S}_n(t^-) \frac{\delta_i dI(X_{k + 1, i} > t)}{I(X_{k + 1, i} > t^-)} \\
& + 
\sum_{i = 1}^n \left\{ \frac{(1 - \delta_i) dI(X_i \leq t)}{\hat{S}_n(X_i)}\right\} \hat{S}_n(t).
\end{align*}

Using the fact that $\sum_{i=1}^{n} \psi_{\hat{S}_{n}, t^-} = 0$, then, $\sum_{i = 1}^n \left\{ \frac{(1 - \delta_i) I(X_i \leq t^-)}{\hat{S}_n(X_i)} - 1 \right\} \hat{S}_n(t^-) = - \sum_{i = 1}^n \delta_i S_{k + 1, 0}(t^- - X_i | H_{K + 1}) - \sum_{i = 1}^n (1 - \delta_i) I(X_i > t^-) \\
= \sum_{i = 1}^n \bigg\{ -\delta_i S_{k+1, 0} (t^- - x_i | H_{k + 1}) - (1 - \delta_i) I(X_i > t^-) \bigg\}.$

We substitute this into our original expression, so,

\begin{align*}
& \sum_{i=1}^{n} \psi_{\hat{S}_{n}, t} = 
\sum_{i = 1}^n \delta_i dS_{k + 1, 0}(t - X_i | H_{K + 1}) + \sum_{i = 1}^n (1 - \delta_i) dI(X_i > t) \\
& + 
\sum_{i = 1}^n \left\{ \frac{(1 - \delta_i) dI(X_i \leq t)}{\hat{S}_n(X_i)}  \right\} \hat{S}_n(t) \\
& + \sum_{i = 1}^n \bigg\{ -\delta_i S_{k+1, 0} (t^- - x_i | H_{k + 1}) - (1 - \delta_i) I(X_i > t^-) \bigg\}\frac{\delta_i dI(X_{k + 1, i} > t)}{I(X_{k + 1, i} > t^-)}.
\end{align*}

We note that the final term can be simplified, as $I(X_{k + 1, i} > t^-) = S_{k + 1, 0}(t^- - X_i)$, so the final term can be rewritten as $\sum_{i = 1}^n \delta_i dI(X_{k + 1, i} > t)$. This results in

\begin{align*}
& \sum_{i=1}^{n} \psi_{\hat{S}_{n}, t} = 
\sum_{i = 1}^n (1 - \delta_i) dI(X_i > t) +
\sum_{i = 1}^n \left\{ \frac{(1 - \delta_i) dI(X_i \leq t)}{\hat{S}_n(X_i)} \right\} \hat{S}_n(t).
\end{align*}

Now, we note that when $\delta_i = 0$, $dI(X_i > t) = -1$ and $\hat{S}(X_i) = \hat{S}(t)$. Otherwise, $dI(X_i > t) = 0$.

Now, we encounter two possible scenarios depending on the value of $\delta$, however we will show that they result in the same conclusion.

\[
\underline{\text{Case 1:}} \ \delta = 1; \quad
\sum_{i=1}^{n} \psi_{\hat{S}_{n}, t} = 0.
\]

\[
\underline{\text{Case 2:}} \ \delta = 0 ;\quad
- \sum_{i = 1}^n (1 - \delta_i) I(X_i = t) +
\sum_{i = 1}^n \left\{ \frac{(1 - \delta_i) I(X_i = t)}{\hat{S}_n(X_i)}  \right\} \hat{S}_n(t) = 0.
\]

\vspace{5mm}

For case 2, this is because if $X_i = t, \hat{S}_n(X_i) = \hat{S}_t$ and $dI(X_i) > t) = -1$. Otherwise, the jump size is 0.

$\Box$

\vspace{5mm}

\textit{\underline{Proof of Proposition 10.}} As \(n \to \infty\), \(\sup_{S \in \Theta_n} \|\Psi_n(S) - \Psi(S)\|_L \to 0\) in probability.

The proof of proposition 10 requires proposition 5, which uses entropy calculations and empirical process theory to show that the tree (and therefore the forest) kernels are Donsker, as well as $\psi$. This has already been proved earlier.

\textbf{Recall: Proposition 5.} The collections of the unnormalized tree and forest kernel functions are Donsker, where the tree kernels $k^{tree}(\cdot)$ are axis-aligned random hyper-rectangles, and the forest kernels $k^{forest}(\cdot)$ are the mean of arbitrarily many tree kernels (n$_{tree}$).

For the proof of proposition 10, we will first show that the class of functions, $\bigg \{ \Psi_{s, t}: S \in \Theta_n, t \in [0, \tau_{lj}]\bigg\}$ and $\bigg \{ \widetilde{\Psi}_{s, t, \widetilde{S}_{k+ 1}}: S \in \Theta_n, \widetilde{S}_{k + 1} \in \Theta_{k + 1}, t \in [0, \tau_{lj}]\bigg\}$ are Donsker. By similar arguments in \cite{cho2023multi}, these two classes are Donsker, since all of their terms belong to a uniformly bounded Donsker class by lemma 9.10 of \cite{kosorok2008introduction}. Therefore, the set of functions is Donsker. Using proposition 5 and the fact that the class of products of bounded Donsker classes is Donsker, $\bigg \{ \Psi_{s, t, \delta_h}: S \in \Theta_n, t\in [0, \tau_{lj}], h \bigg \}$, $\bigg \{\Psi_{s, t, k_h}: S \in \Theta_n, t\in [0, \tau_{lj}], h \in \mathcal{H} \bigg \}$, and $\bigg \{ \widetilde{\Psi}_{s, t, \widetilde{S}_{k + 1}k_h}: S \in \Theta_n, \widetilde{S}_{k + 1} \in \Theta_{k + 1}, t \in [0, \tau_{lj}, h \in \mathcal{H}]\bigg\}$ are Donsker. \\

By assumption 1 (terminal node size, polynomial), we assume that the minimum size of the terminal nodes ($n_{\text{min}}$) grow at rate $n_{\text{min}} \asymp n^\beta$, $\frac{1}{2} < \beta < 1$; so $\sup_h \mathbb{P}_nk_h \asymp n^{\beta - 1}$.

\vspace{5mm}

Then,

\[
\sup_{S \in \Theta_n, \widetilde{S}_{k + 1} \in \Theta_{k + 1}} \|\Psi_n(S) - \Psi(S)\|_L = \sup_{S \in \Theta_n, \widetilde{S}_{k + 1} \in \Theta_{k + 1}} \| \frac{\mathbb{P}_n \widetilde{\psi}_{S,t} k_h }{\mathbb{P}_n k_h} - \frac{P \psi_{S,t} \delta_h}{P\delta_h}  \|_\mathbb{L}
\]
\[
= \sup_{S \in \Theta_n, \widetilde{S}_{k + 1} \in \Theta_{k + 1}} \|  \frac{\mathbb{P}_n \widetilde{\psi}_{S,t} k_h }{\mathbb{P}_n k_h} - \frac{\mathbb{P}_n \psi_{s, t} k_h}{\mathbb{P}_nk_h} + \frac{\mathbb{P}_n \psi_{s, t} k_h}{\mathbb{P}_nk_h} - \frac{P\psi_{s, k}k_h}{Pk_h} + \frac{P\psi_{s, k}k_h}{Pk_h} - \frac{P \psi_{S,t} \delta_h}{P\delta_h}\|_\mathbb{L}.
\]

\vspace{5mm}

By triangle inequality, 

\begin{align*}
&\leq \sup_{S \in \Theta_n, \widetilde{S}_{k + 1} \in \Theta_{k + 1}} 
\left| \frac{\mathbb{P}_n \widetilde{\psi}_{S,t} k_h}{\mathbb{P}_n k_h} - \frac{\mathbb{P}_n \psi_{S,t} k_h}{\mathbb{P}_n k_h} \right| && \text{(A)} \\
&+ \sup_{S \in \Theta_n, \widetilde{S}_{k + 1} \in \Theta_{k + 1}} 
\left| \frac{\mathbb{P}_n \psi_{S,t} k_h}{\mathbb{P}_n k_h} - \frac{P \psi_{S,t} k_h}{P k_h} \right| && \text{(B)} \\
&+ \sup_{S \in \Theta_n, \widetilde{S}_{k + 1} \in \Theta_{k + 1}} 
\left| \frac{P \psi_{S,t} k_h}{P k_h} - \frac{P \psi_{S,t} \delta_h}{P \delta_h} \right| && \text{(C)} \\
= o_p(1).
\end{align*}

We will prove that each term (A), (B), (C) are $o_p(1)$, which means that added together they are $o_p(1)$. 

\vspace{5mm}
We start with 
\[
(A)= 
\sup_{S \in \Theta_n, \widetilde{S}_{k + 1} \in \Theta_{k + 1}} 
\left| \frac{\mathbb{P}_n \widetilde{\psi}_{S,t} k_h}{\mathbb{P}_n k_h} - \frac{\mathbb{P}_n \psi_{S,t} k_h}{\mathbb{P}_n k_h} \right| 
\]

\begin{align*}
= &\sup_{S \in \Theta_n, \widetilde{S}_{k + 1} \in \Theta_{k + 1}} 
\Bigg| \frac{\mathbb{P}_n \left\{ \delta \widetilde{S}_{k + 1}(t - x | H_{k + 1}) + (1 - \delta) 
    \left( \frac{S(t | H)}{S(X | H)} \land 1 \right) - S(t | H) \right\}  k_h}{\mathbb{P}_n k_h} \\
&\quad - \frac{\mathbb{P}_n \left\{ \delta S_{k + 1, 0}(t - x | H_{k + 1}) + (1 - \delta) 
    \left( \frac{S(t | H)}{S(X | H)} \land 1 \right) - S(t | H) \right\}  k_h}{\mathbb{P}_n k_h} \Bigg|
\end{align*}
\[
=
\sup_{S \in \Theta_n, \widetilde{S}_{k + 1} \in \Theta_{k + 1}} \bigg| \frac{\mathbb{P} \delta \{\widetilde{S}_{k + 1} (t - X | H_{k + 1}) - S_{k + 1, 0} (t - X | H_{k + 1}) \} k_h}{\mathbb{P}_n k_h}  
\bigg|
\]
\[
\leq 
\sup_{S \in \Theta_n, \widetilde{S}_{k + 1} \in \Theta_{k + 1}} \bigg| \frac{\mathbb{P}_n \delta k_h}{\mathbb{P}_nk_h} \bigg| \cdot
\sup_{S \in \Theta_n, \widetilde{S}_{k + 1} \in \Theta_{k + 1}} \bigg| \widetilde{S}_{k + 1}(t - X | H_{k + 1}) - S_{k + 1, 0}(t - X | H_{k + 1}) \bigg|,
\]
\vspace{5mm}

where the first term is $\leq 1$, and the second term is $o_p(1)$. Therefore, (A) is $o_p(1)$.

\[
(B)= 
\sup_{S \in \Theta_n, \widetilde{S}_{k + 1} \in \Theta_{k + 1}} 
\left| \frac{\mathbb{P}_n \psi_{S,t} k_h}{\mathbb{P}_n k_h} - \frac{P \psi_{S,t} k_h}{P k_h} \right| =
\sup_{S \in \Theta_n, \widetilde{S}_{k + 1} \in \Theta_{k + 1}} \bigg| \frac{\mathbb{P}_n \psi_{S,t} k_h}{\mathbb{P}_n k_h} - \frac{P\psi_{s, k} k_h}{\mathbb{P}k_h} + \frac{P\psi_{s, k} k_h}{\mathbb{P}k_h} - \frac{P \psi_{S,t} k_h}{P k_h} \bigg|
\]

\begin{align*}
= &\sup_{S \in \Theta_n, \widetilde{S}_{k + 1} \in \Theta_{k + 1}} 
\left| \frac{\mathbb{P}_n \psi_{S,t} k_h - P \psi_{S,k} k_h}{\mathbb{P}_n k_h} \right| && \text{(B1)} \\
&+ \sup_{S \in \Theta_n, \widetilde{S}_{k + 1} \in \Theta_{k + 1}} 
\left| P \psi_{S,k} k_h \left( \frac{1}{\mathbb{P}_n k_h} - \frac{1}{P k_h} \right) \right| && \text{(B2)} \\
= o_p(1).
\end{align*}

We look at these terms individually, and prove that both terms are $o_p(1)$, therefore (B) is $o_p(1)$.

(B1) = Using the fact that $\sup_h \mathbb{P}_nk_h \asymp n^{\beta - 1}$,
\[
(B1) \leq \frac{\mathcal{O}_p(n^{-\frac{1}{2}})}{n^{\beta - 1}} = \mathcal{O}_p(\frac{1}{n^{\beta - \frac{1}{2}}}) = o_p(1).
\]

(B2)= 
\[
\sup_{S \in \Theta_n, \widetilde{S}_{k + 1} \in \Theta_{k + 1}} 
\left| P \psi_{S,k} k_h \left( \frac{1}{\mathbb{P}_n k_h} - \frac{1}{P k_h} \right) \right| = \sup_{S \in \Theta_n, \widetilde{S}_{k + 1} \in \Theta_{k + 1}} \bigg| P(\psi_{s, k}k_h) \{ \frac{Pk_h - \mathbb{P}_nk_h}{Pk_h \mathbb{P}_nk_h}\} \bigg| 
\]
\[
= \sup_{S \in \Theta_n, \widetilde{S}_{k + 1} \in \Theta_{k + 1}} \bigg| \frac{P\psi_{s, k}k_h}{Pk_h} \cdot \{ \frac{-(\mathbb{P}_n - P)k_h}{\mathbb{P}_nk_h} \} \bigg| 
\leq
\frac{\mathcal{O}_p(n^{-\frac{1}{2}})}{n^{\beta - 1}} = o_p(1).
\]

\[
(C)=
\sup_{S \in \Theta_n, \widetilde{S}_{k + 1} \in \Theta_{k + 1}} 
\bigg| \frac{P \psi_{S,t} k_h}{P k_h} - \frac{P \psi_{S,t} \delta_h}{P \delta_h} \bigg| = \sup_{S \in \Theta_n, \widetilde{S}_{k + 1} \in \Theta_{k + 1}} \bigg|P_{\cdot|k_h \psi_{s,t}} -  P_{\cdot|h \psi_{s,t}} \bigg|.
\]
\\
We first note that $S(t | k_h) = S(t | h)$ for all $S \in \Theta_n$. Then, for all $S \in \Theta_n, t \in [0, \tau_{lj}]$ and $h \in \mathcal{H}$. Therefore, 

\[
\begin{aligned}
(C) = \bigg| & S_0(t | k_h) \{1 - G(t | k_h)\} - S(t | k_h) 
+ S(t | k_h) \int_0^t \frac{S_0(u | k_h)}{S(u | k_h)} dG(u | k_h) \\
& - S_0(t | h) \{1 - G(t | h)\} + S(t | h) 
- S(t | h) \int_0^t \frac{S_0(u | h)}{S(u | h)} dG(u | h)
\bigg|
\end{aligned}
\]
\[
= \bigg| S_0(t |k_h) \{1 - G(t | k_h) \} - S_0(t | h) \{1 - G(t|h) \} + S(t | h) \bigg[ \int_0^t \frac{S_0(u | k_h)}{S(u|k_h)}dG(u | k_h) - \int_0^t \frac{S_0(u | h)}{S(u | h)} dG(u | h) \bigg]
\bigg|.
\]

\vspace{5mm}

Now, by the triangle inequality,

\begin{align*}
\leq & \bigg| S_0(t | k_h) \{1 - G(t | k_h)\} - S_0(t | h) \{1 - G(t | h)\} \bigg| \\
&+ \bigg| S(t | h) \int_0^t\bigg[  \frac{S_0(u | k_h)}{S(u|k_h)}dG(u | k_h) - 
\frac{S_0(u | h)}{S(u | h)}dG(u | h) + \frac{S_0(u | h)}{S(u | h)}dG(u | h) - \frac{S_0(u | h)}{S(u | h)} dG(u | h)\bigg]\bigg| \\
 = & \bigg| S_0(t | k_h) \{1 - G(t | k_h)\} - S_0(t | h) \{1 - G(t | h)\} \bigg| \\
&+ \bigg| S(t | h)  \int_0^t \bigg\{ \frac{S_0(u | k_h)}{S(u|h)}\bigg[ dG(u | k_h) - dG(u | h) \bigg] + 
dG(u | h) \bigg[\frac{S_0(u | k_h)}{S(u | h)} - \frac{S_0(u | h)}{S(u | h)} \bigg] \bigg\} \bigg|
\end{align*}

\vspace{5mm}

By triangle inequality on the second term, we can obtain 

\begin{align*}
\leq & \bigg| S_0(t | k_h) \{1 - G(t | k_h)\} - S_0(t | h) \{1 - G(t | h)\} \bigg| && \text{(C1)} \\
&+ \bigg| \int_0^t \frac{S_0(u | k_h)}{S(u | h)} \bigg[ dG(u | k_h) - dG(u | h) \bigg] \cdot S(t | h) \bigg| && \text{(C2)} \\
&+ \bigg| \int_0^t dG(u | h) \bigg[ \frac{S_0(u | k_h) - S_0(u | h)}{S(u | h)} \cdot S(t | h) \bigg] \bigg|,  && \text{(C3)}
\end{align*}

\vspace{5mm}

where we will  further bound each of these terms.

(C1)=
\begin{align*}
\bigg| S_0(t | k_h) \{1 - G(t | k_h)\} - S_0(t | h) \{1 - G(t | h)\} \bigg| \\
&= \bigg| S_0(t | k_h) \{1 - G(t | k_h)\} - S_0(t | h) \{1 - G(t | h)\} \\
&\quad + S_0(t | k_h) \{1 - G(t | h)\} - S_0(t | k_h) \{1 - G(t | h)\}  \bigg| \\
&= \bigg| S_0(t | k_h) \bigg[ \{1 - G(t | k_h)\} - \{1 - G(t | h)\} \bigg] \\
&\quad + \{1 - G(t | h)\} \bigg[ S_0(t | k_h) - S_0(t | h) \bigg] \bigg|.
\end{align*}

Through triangle inequality,

\begin{align*}
\leq 
\bigg| S_0(t | k_h) \bigg[ G(t | h) - G(t | k_h) \bigg]
\bigg| +  \bigg| \{1 - G(t | h)\} \bigg[ S_0(t | k_h) - S_0(t | h) \bigg] \bigg| \\
=
S_0(t | k_h) \bigg| G(t | h) - G(t | k_h)
\bigg| + \{1 - G(t | h)\} \bigg|  S_0(t | k_h) - S_0(t | h)  \bigg|,
\end{align*}

\vspace{5mm}

where the equality is because $0 \leq S_0(t|k_h) \leq 1$ and $0 \leq 1 - G(t | h) \leq 1$.\\

Now, using assumption 3 (lipschitz continuous), which holds for all $h_1, h_2\in \mathcal{H}, t \in [0, \tau_{lj} - B_k]$,

\begin{align*}
\leq S_0(t | k_h) L_G \|h - h' \|_\infty + \{1 - G(t | h)\} L_S \|h - h' \|_\infty  \\
\leq L_G \|h - h' \|_\infty + L_S \|h - h' \|_\infty \\
\leq L_G \sup_{h' \in \mathcal{H}:k_h(h') > 0}\|h - h' \|_\infty + L_S \sup_{h' \in \mathcal{H}:k_h(h') > 0}\|h - h' \|_\infty,
\end{align*}

where the first to second line inequality is because $0 \leq S_0(t | k_h) \leq 1$ and $0 \leq 1 - G(t | h) \leq 1$. \\

(C2)=
\begin{align*}
\bigg| \int_0^t \frac{S_0(u | k_h)}{S(u | h)} \bigg[ dG(u | k_h) - dG(u | h) \bigg] \cdot S(t | h) \bigg| \\
& \leq \int_0^t S_0(u | k_h) \frac{S(t | h)}{S(u | h)} \bigg| dG(u | k_h) - dG(u | h) \bigg| \\
& \leq \int_0^t \bigg| dG(u | k_h) - dG(u | h) \bigg|,
\end{align*}

\vspace{5mm}

where the first to second line inequality stems from bringing the integral outside the absolute value, the second to third line inequality is becaue $\frac{S(t | h)}{S(u | h)} \leq 1$ and $0 < S_0(u | k_h) \leq 1$. We also note that $\bigg| G_k(t | h_1) - G_k(t | h_2) \bigg|\leq L_G \| h_1 - h_2 \|_\infty $ by assumption, so $\bigg| dG_k(t | h_1) -dG_k(t | h_2) \bigg| \leq 2L_G \| h_1 - h_2 \|_\infty dt $, so

\begin{align*}
(C2) \leq \int_0^t 2 L_G \| h - h' \|_\infty du \\
\leq \int_0^t 2 L_G \sup_{h' \in \mathcal{H}:k_h(h') > 0}\|h - h' \|_\infty du \\
\leq 2t L_G \sup_{h' \in \mathcal{H}:k_h(h') > 0}\|h - h' \|_\infty.
\end{align*}

(C3)= 
\begin{align*}
\bigg| \int_0^t dG(u | h) \bigg[ \frac{S_0(u | k_h) - S_0(u | h)}{S(u | h)} \cdot S(t | h) \bigg] \bigg| = \bigg| \int_0^t \frac{S(t | h)}{S(u | h)} \bigg[S_0(u | k_h) - S_0(u | h) \bigg] dG(u | h) \bigg| \\
\leq \bigg| \int_0^t \bigg[ S_0(u | k_h) - S_0(u | h) \bigg] dG(u | h) \bigg| \\
\leq \sup_t \sup_h \bigg| S_0(t | k_h) - S_0(t | h) \bigg| \cdot \int_0^t dG(u |h),
\end{align*}

where the first inequality follows because $0 \leq \frac{S(t | h)}{S(u | h)} \leq 1$. \\

Using the fact that $\int_0^t dG(u |h) = G(t | h)$, and $0 < G(t | h) \leq 1$, we can then get

\begin{align*}
\leq \sup_t \sup_h \bigg| S_0(t | k_h) - S_0(t | h) \bigg| \\
\leq L_S \| h - h' \|_\infty \\
\leq L_S  \sup_{h' \in \mathcal{H}:k_h(h') > 0} \| h - h' \|_\infty,
\end{align*}

where the first to the second line inequality drops $\sup_t$ since the term is no longer a function of $t$.

Putting all the terms for (C) together, we get,

\begin{align*}
\sup_{S \in \Theta_n, \widetilde{S}_{k + 1} \in \Theta_{k + 1}} \bigg|P_{\cdot|k_h \psi_{s,t}} -  P_{\cdot|h \psi_{s,t}} \bigg| \\
& \leq L_G \sup_{h' \in \mathcal{H}:k_h(h') > 0}\|h - h' \|_\infty + L_S \sup_{h' \in \mathcal{H}:k_h(h') > 0}\|h - h' \|_\infty  \\
&+ 2t L_G \sup_{h' \in \mathcal{H}:k_h(h') > 0}\|h - h' \|_\infty + L_S  \sup_{h' \in \mathcal{H}:k_h(h') > 0} \| h - h' \|_\infty \\
&= \bigg\{(2t + 1) L_G + 2 L_S \bigg\}  \sup_{h' \in \mathcal{H}:k_h(h') > 0}\|h - h' \|_\infty \\
& \leq \bigg\{(2\tau_{lj} + 1) L_G + 2 L_S \bigg\}  \sup_{h' \in \mathcal{H}:k_h(h') > 0}\|h - h' \|_\infty,
\end{align*}

where the last inequality comes from the fact that $t \leq \tau_{lj}$. \\

The remainder of the proof shows $\sup_{h' \in \mathcal{H}:k_h(h') > 0}\|h - h' \|_\infty = o_p(1)$, which is done in the arguments \cite{cho2023multi} using \cite{meinshausen2006quantile}.

$\Box$

%% file: Supplement_Proofs/value_supplement.tex
We first introduce some notation for the intermediate visit Q and V functions, also known as "state-action-value" and "state-value" functions in the reinforcement learning field. Let $Q^\pi(H_1, a) = \phi \{S^\pi (\cdot | H_1, A_1 = a) \}$ and $\mathcal{V}^\pi(H_1) = \phi\{ S^\pi (\cdot | H_1, A_1 = \pi_1(H_1))) \}$. We modify the approach taken by \cite{cho2023multi}, which used modified versions of Lemma 1 and 2 from \cite{murphy2005generalization}. The difference in our approach is that the patient's first visit will account for their entire optimized survival trajectory, assuming the patient receives the optimal in the future. Therefore, we can examine the patient's trajectory at the first visit only. However, in the original \cite{cho2023multi} paper, each visit's decision is made on a separate model; thus, the need for the sum across all visits $k = 1, ... , K$. 

\vspace{5mm}

\setlength{\parindent}{0pt}

\textbf{Lemma 3.} For treatment regimes $\{\tilde{\pi} \}$ and $\{ \pi \}$, 

\[
\mathcal{V}(\tilde{\pi}) - \mathcal{V}(\pi) = - \mathbb{E}^\pi \bigg[ \Delta^{\tilde{\pi}}(H_1, A_1)\bigg],
\]

where $\Delta^{\tilde{\pi}}(H_1, A_1) = \mathcal{Q}^{\tilde{\pi}}(H_1, A_1) - \mathcal{V}^{\tilde{\pi}}(H_1)$ is the advantage of treatment $A_1 = a$ under policy $\pi$ over the treatment $\tilde{\pi}(H_1)$ prescribed by $\tilde{\pi}$ at the first visit.

\vspace{5mm}

\underline{Proof of Lemma 3.} \\

We define the value of the policy as the expected value of the area under the curve at the first visit, since at the first visit, the rest of the patient trajectory at subsequent visits has already been maximized. We define the Q function under $\pi$ as the expected survival outcome starting from $H_1$, taking action $a$, and following policy $\pi$ thereafter.

\[
\mathcal{V}(\pi) = \mathbb{E} \bigg[ \mathcal{V}^\pi (H_1) \bigg] =  \mathbb{E} \bigg[ \mathcal{Q}^{\pi} (H_1, \pi(H_1)) \bigg] = \mathbb{E} \bigg[ \phi\{ S^\pi (\cdot | H_1, \pi(H_1)) \}\bigg].
\]

\[
\mathcal{V}(\tilde{\pi}) = \mathbb{E} \bigg[ \mathcal{V}^{\tilde{\pi}} (H_1) \bigg] = \mathbb{E} \bigg[ \mathcal{Q}^{\tilde{\pi}} (H_1, \tilde{\pi}(H_1)) \bigg] = \mathbb{E} \bigg[ \phi\{ S^{\tilde{\pi}} (\cdot | H_1, \tilde{\pi}(H_1)) \}\bigg].
\]

Then,

\[
\mathcal{V}(\tilde{\pi}) - \mathcal{V}(\pi) = 
\mathbb{E} \bigg[ \mathcal{V}^{\tilde{\pi}} (H_1) - \mathcal{V}^{\pi} (H_1) \bigg]
\]
\[
= \mathbb{E} \bigg[ \phi\{ S^{\tilde{\pi}} (\cdot \mid H_1, \tilde{\pi}(H_1)) \}\bigg] - 
 \mathbb{E} \bigg[ \phi\{ S^\pi (\cdot \mid H_1, \pi(H_1)) \}\bigg]. \eqno{(1)}
\]

 \vspace{5mm}

 We note that $\mathcal{V}^{\pi}(H_1) = \phi \{ S^\pi (\cdot | H_1, \pi(H_1)) \} = \mathcal{Q}^\pi (H_1, A_1)$, where $A_1 = \pi(H_1)$. We can rewrite this by adding and subtracting the same term, resulting in

 \[
 \mathcal{V}^\pi(H_1) = \phi \{S^\pi (\cdot | H_1, \pi(H_1)) \} - \phi \{S^{\tilde{\pi}} (\cdot | H_1, A_1) \} + \phi \{S^{\tilde{\pi}} (\cdot | H_1, A_1) \}
 \]
 \[
 = \mathcal{Q}^{\tilde{\pi}}(H_1, A_1) + \bigg[ \phi \{S^\pi (\cdot | H_1, \pi(H_1)) \} - \phi \{S^{\tilde{\pi}} (\cdot | H_1, A_1) \} \bigg].
 \]

 \vspace{5mm}

 We then note that since the future policies are optimal after visit $k = 1$, the survival probabilities from $k = 2$ and onwards are the same for both policies. Therefore, $S^\pi( \cdot | H_1, \pi(H_1)) = S^{\tilde{\pi}} (\cdot | H_1, \pi(H_1))$ and $S^{\tilde{\pi}}( \cdot | H_1, \tilde{\pi}(H_1)) = S^\pi (\cdot | H_1, \tilde{\pi}(H_1))$.

 \vspace{5mm}

 Thus, for $A_1 = \pi(H_1)$ since we are following policy $\pi$, we can rewrite as
 \[
\mathcal{V}^\pi(H_1) = \mathcal{Q}^{\tilde{\pi}}(H_1, A_1 = \pi(H_1)).
 \]

 \vspace{5mm}

 Going back to (1), we substitute the above expression, resulting in 

 \[
 \mathcal{V}(\tilde{\pi}) - \mathcal{V}(\pi) = \mathbb{E} \bigg[ \mathcal{V}^{\tilde{\pi}}(H_1) - \mathcal{Q}^{\tilde{\pi}}(H_1, A_1 = \pi(H_1)) \bigg]
 \]

\[ =
- \mathbb{E} \bigg[ \Delta^{\tilde{\pi}}(H_1, A_1 = \pi(H_1)) \bigg],
\]

\vspace{5mm}

where $\Delta^{\tilde{\pi}}(H_1, A_1) = \mathcal{Q}^{\tilde{\pi}}(H_1, A_1) - \mathcal{V}^{\tilde{\pi}}(H_1)$ measures how much better or worse taking $A_1$ under $\pi$ is, compared to the action prescribed by policy $\tilde{\pi}$ at $H_1$. This is then written as

\vspace{5mm}

\[
= - \mathbb{E}^\pi \bigg[ \Delta^{\tilde{\pi}}(H_1, A_1) \bigg],
\]

\vspace{5mm}

where $\mathbb{E}^\pi$ is the expectation with respect to a measure $dP^\pi(H_1, A_1)$, meaning we take the expectation over the randomness induced by policy $\pi$. $\Box$

\vspace{5mm}

\textbf{Lemma 4.} For all functions $\hat{\mathcal{Q}}$ satisfying $\hat{\pi}(H_1) \in argmax_a \hat{\mathcal{Q}}(H_1,a)$, we have

\[
\mathcal{V}(\pi^*) - \mathcal{V}(\hat{\pi}) \leq 2 \zeta^{-1} \sqrt{ \mathbb{E}  \left[ \left( Q^{\pi^*}(H_1, A_1) - \hat{Q}(H_1, A_1) \right)^2 \right] },
\]

where $\pi^*$ is the optimal decision rule and $\hat{\pi}$ is the estimated decision rule. 

\vspace{5mm}

\underline{Proof of Lemma 4.} \\

Using Lemma 3, 

\[
\mathcal{V}(\pi^*) - \mathcal{V}(\hat{\pi}) = - \mathbb{E}^{\hat{\pi}} \bigg[ \Delta^{\pi^*} (H_1, A_1) \bigg], \eqno{(2)}
\]

\vspace{5mm}

where we define the advantage function $\Delta^{\pi^*}(H_1, A_1) = \mathcal{Q}^{\pi^*}(H_1, A_1) - \mathcal{V}^{\pi^*}(H_1)$, and $A_1 = \hat{\pi}(H_1)$, where we are evaluating how $\hat{\pi}$'s action impacts the expected value as compared to following the policy prescribed by $\pi^*$.

\vspace{5mm}

Since $\mathcal{V}^{\pi^*}(H_1) = \mathcal{Q}^{\pi^*}(H_1, \pi^*(H_1))$ and $A_1 = \hat{\pi}$, we can rewrite the advantage function as 

\[
\Delta^{\pi^*}(H_1, A_1) = \mathcal{Q}^{\pi^*}(H_1,\hat{\pi}(H_1))- \mathcal{Q}^{\pi^*}(H_1, \pi^*(H_1))
\]

\begin{align*}
= &\left\{ \mathcal{Q}^{\pi^*}(H_1, \hat{\pi}(H_1)) - \hat{\mathcal{Q}}(H_1, \hat{\pi}(H_1)) \right\} \tag{A} \\
&+ \left\{ \hat{\mathcal{Q}}(H_1, \hat{\pi}(H_1)) - \hat{\mathcal{Q}}(H_1, \pi^*(H_1)) \right\} \tag{B} \\
&+ \left\{ \hat{\mathcal{Q}}(H_1, \pi^*(H_1)) - \mathcal{Q}^{\pi^*}(H_1, \pi^*(H_1)) \right\}. \tag{C}
\end{align*}

\vspace{5mm}

We note that $\hat{\pi}(H_1) = argmax_a \hat{\mathcal{Q}}(H_1, a)$, then (B) $\geq 0$. Therefore,

\vspace{5mm}

\[
\Delta^{\hat{\pi}}(H_1) \leq
 \bigg| \mathcal{Q}^{\pi^*}(H_1, \hat{\pi}(H_1)) - \hat{\mathcal{Q}}(H_1, \hat{\pi}(H_1)) \bigg| + \bigg| \hat{\mathcal{Q}}(H_1, \pi^*(H_1)) - \mathcal{Q}^{\pi^*}(H_1, \pi^*(H_1)) \bigg|
\]

\vspace{5mm}
\[
\leq max_a \bigg| \mathcal{Q}^{\pi^*}(H_1, a) - \hat{\mathcal{Q}}(H_1, a) \bigg| + max_a \bigg| \hat{\mathcal{Q}}(H_1, a) - \mathcal{Q}^{\pi^*}(H_1, a) \bigg|
\]

\[
= 2 max_a \bigg| \mathcal{Q}^{\pi^*}(H_1, a) - \hat{\mathcal{Q}}(H_1, a) \bigg|.
\]

Therefore, using (2), 

\[
\mathcal{V}(\pi^*) - \mathcal{V}(\hat{\pi}) \leq \mathbb{E}^{\hat{\pi}} \bigg[ 2 max_a \bigg| \mathcal{Q}^{\pi^*}(H_1, a) - \hat{\mathcal{Q}}(H_1, a) \bigg|\bigg].
\]

\vspace{5mm}

Using Hölder's inequality, and the covariate density bound in Assumption 4,

\[
\leq 2 \zeta^{-1} \sqrt{ \mathbb{E}^{\hat{\pi}} \bigg[ max_a \bigg| \mathcal{Q}^{\pi^*}(H_1, a) - \hat{\mathcal{Q}}(H_1, a) \bigg| \bigg]^2}
\]

\[
\leq 2 \zeta^{-1} \sqrt{ \mathbb{E}^{\hat{\pi}} \bigg[ \sum_{a \in \mathcal{A}} \bigg| \mathcal{Q}^{\pi^*}(H_1, a) - \hat{\mathcal{Q}}(H_1, a) \bigg| P(A_1 = a | H_1) \bigg]^2}
\]

\[
\leq 2 \zeta^{-1/2} \sqrt{ \mathbb{E} \bigg[ \mathcal{Q}^{\pi^*}(H_1, A_1) - \hat{\mathcal{Q}}(H_1, A_1) \bigg]^2}.
\]

\vspace{5mm}

Next, we will bound the tail quantity in the expression above.

\[
\mathbb{E} \left[ \{ \mathcal{Q}^{\pi^*}(H_1, A_1) - \hat{\mathcal{Q}}(H_1, A_1) \}^2 \right] \leq 
\sup_{h_1, a_1} \{ \mathcal{Q}^{\pi^*}(h_1, a_1) - \hat{\mathcal{Q}}(h_1, a_1) \}^2
\]

\[
= \sup_{h_1, a_1} \bigg[ \phi \{ S^*_1( \cdot | h_1, a_1; S_1^*) - \hat{S}_1(\cdot | h_1, a_1; \hat{S}_1) \}\bigg]^2
\]

\[
\leq c(\phi)^2 \sup_{h_1, a_1} \bigg[ \phi \{ S_1^*( \cdot | h_1, a_1; S^*_1) - \hat{S}_1(\cdot | h_1, a_1; \hat{S}_1) \}\bigg]^2,
\]

where the inside $\hat{S}_1$ are survival probabilities obtained from the two-step prediction process used for refitting. The equality comes from the fact that $\phi$ is a linear operator. For the inequality, $c(\phi) = \tau$ since the Q-function is bounded by $[0, \tau]$.

\vspace{5mm}

Substituting this back into the original context,

\[
\mathcal{V}(\pi^*) - \mathcal{V}(\hat{\pi}) \leq 2 \zeta^{-1/2} c(\phi) \sup_{h_1, a_1} \bigg[ \phi \{ S^*_1( \cdot | h_1, a_1; S^*_1) - \hat{S}_1(\cdot | h_1, a_1; \hat{S}_1) \}\bigg].
\]

For the same reasons as \cite{cho2023multi}, we cannot directly apply Theorem 2, which states that $\sup_{t \in [0, \tau_{lj}], h_k \in \mathcal{H}} |\hat{S}_k (t \mid h_k) - S_k (t \mid h_k)| \to 0$, as the optimized survival estimator $\hat{S}$ may not be consistent for the true optimal survival curve $S^*$. So, we need to further break this down into 

\begin{align*}
\mathcal{V}(\pi^*) - \mathcal{V}(\hat{\pi}) 
&\leq  2 \zeta^{-1/2} c(\phi) \sup_{h_1, a_1} \bigg[ 
    \phi\{ S_1^*(\cdot | h_1, a_1; S^*_1)\} 
    - \phi\{ S_1^*( \cdot | h_1, a_1; \hat{S}_1) \} \\
&\quad + \phi \{S_1^*( \cdot | h_1, a_1; \hat{S}_1) \} 
    - \phi \{\hat{S}_1(\cdot | h_1, a_1; \hat{S}_1) \} 
\bigg].
\end{align*}

\vspace{5mm}

By triangle inequality,

\begin{align*}
&\leq 2 \zeta^{-1/2} c(\phi) \sup_{h_1, a_1} 
\left[ \phi \left\{ S_1^*(\cdot \mid h_1, a_1; \hat{S}_1) - \hat{S}_1(\cdot \mid h_1, a_1; \hat{S}_1) \right\} \right]  \\
&\quad + 2 \zeta^{-1/2} c(\phi) \sup_{h_1, a_1} 
\left[ \phi \left\{ S_1^*(\cdot \mid h_1, a_1; S^*_1) - S_1^*(\cdot \mid h_1, a_1; \hat{S}_1) \right\} \right] \\
&\leq 2 \zeta^{-1/2} c(\phi)^2 \sup_{h_1, a_1, t \in [0, \tau_{lJ}]} | S_1^*( \cdot | h_1, a_1; \hat{S}_1) - \hat{S}_1( \cdot | h_1, a_1; \hat{S}_1) | \tag{A} \\
&\quad + 2 \zeta^{-1/2} c(\phi)^2 \sup_{h_1, a_1, t \in [0, \tau_{lJ}]} |S_1^* (\cdot | h_1, a_1; S^*_1) - S_1^*(\cdot | h_1, a_1; \hat{S}_1) |. \tag{B}
\end{align*}

\vspace{5mm}

For term (A), we can prove this goes to $0$ in probability by introducing a new lemma (lemma 5), similar to Theorem 1 and Theorem 2.

\vspace{5mm}

\textbf{Lemma 5.} Assume the same set of conditions as in Theorem 2, and that J is the final iteration of the random forest. Then
\[
\sup_{t \in [0, \tau_{lJ}], h_1 \in \mathcal{H}_1, a_1 \in A_1} 
\left| \hat{S}_1(t \mid h_1, a_1; \hat{S}_{1}) - S_1(t \mid h_1, a_1; \hat{S}_{1}) \right| \to 0, \text{ in probability as \( n \to \infty \).}
\]

\vspace{5mm}

\underline{Proof of Lemma 5.} \\
The proof of Lemma 5 is similar to the proof of Theorem 2, In proving Lemma 6, term (A) in the proof of proposition 10 being omitted, with all else being the same.

\vspace{5mm}

For term (B)
\[\sup_{h_1, a_1, t \in [0, \tau_{lJ}]} |S_1^* (t | h_1, a_1; \hat{S}_1) - S_1^*(t | h_1, a_1; S^*_1) |\]
\[
\leq sup_{t, h_1} | Pr(V_1 + L_2 \geq t; L_2 \sim \hat{S}_1( \cdot | h_2) - Pr(V1 + L_2 \geq t; L_2 \sim S^*_1( \cdot | h_2) |
\]
\[
= sup_{t, h_1} | \int I(v_1 + l_2 \geq t) \{d\hat{S}_1(l_2; h_2) - dS^*_1(l_2; h_2) \}dP(h_2) |
\]
\[
\leq sup_{t, h_1} | \hat{S}_1(t, h_1) - S_1^*(t, h_1) |, 
\]

where $L_2$ is the remaining life at visit $2$ and $V_1$ is the uncensored first visit length. 

We then combine terms (A) and (B) to get $4 \zeta^{-1/2} c(\phi)^2 \sup_{h_1, a_1, t \in [0, \tau_{lJ}]} | S_1^*( \cdot | h_1, a_1; \hat{S}_1) - \hat{S}_1( \cdot | h_1, a_1; \hat{S}_1) | $, which goes to 0 in probability from Lemma 5. Note that the use of the $S_1$ in the proof is because the survival curve applies to all time points, not just the first.

$\Box$

%% file: Supplement_Proofs/RDA_details.tex
We derive an optimal dynamic treatment regime following an $80-20$ 10-fold cross validation scheme for value function estimation. We also assume that causal assumptions 8-10 (stable unit treatment value assumption, sequential ignorability, positivity) hold true. 

The training set is used to fit the forest, which outputs a DTR estimator. Using this, predictions can be made on the test set as to what the optimal DTR would be for the test set patients. A propensity score for each observation (each patient and each stage) can then be estimated using a logistic regression model using the same sets of covariates used in the DTR estimation, with treatment being the outcome. Similarly, the censoring probabilities can be estimated using random survival forests. A weight for each observation is then calculated based on if a patient truly received their estimated optimal treatment, shown mathematically as 

\[
W_i(\pi) = \frac{\prod_{k=1}^{3} 1\left(\pi_k(\mathbf{H}_{k,i}) = A_{k,i}\right)\delta_i}{\hat{p}(A_{1,i}) \hat{p}(A_{2,i} \mid \mathbf{H}_{2,i}) \hat{p}(A_{3,i} \mid \mathbf{H}_{3,i}) \prod_{k=1}^{3} \hat{p}(T_{k,i} < C_{k,i} \mid z_i)}.
\]

\vspace{5mm}

The value of the regime is then estimated using the following,

\[
\hat{\phi}(S^\pi) = \frac{\sum_{i \in \text{test set}} \left( T_i \wedge \tau \right) W_i(\pi)}{\sum_{i \in \text{test set}} W_i(\pi)}.
\]

The main manuscript presents results of the analysis using a 2 strata example. Here, we will illustrate the results of the same analysis using a single strata using the same cross-validation approach. 

\begin{figure}[H]
    \centering
    \includegraphics[width=0.65\linewidth]{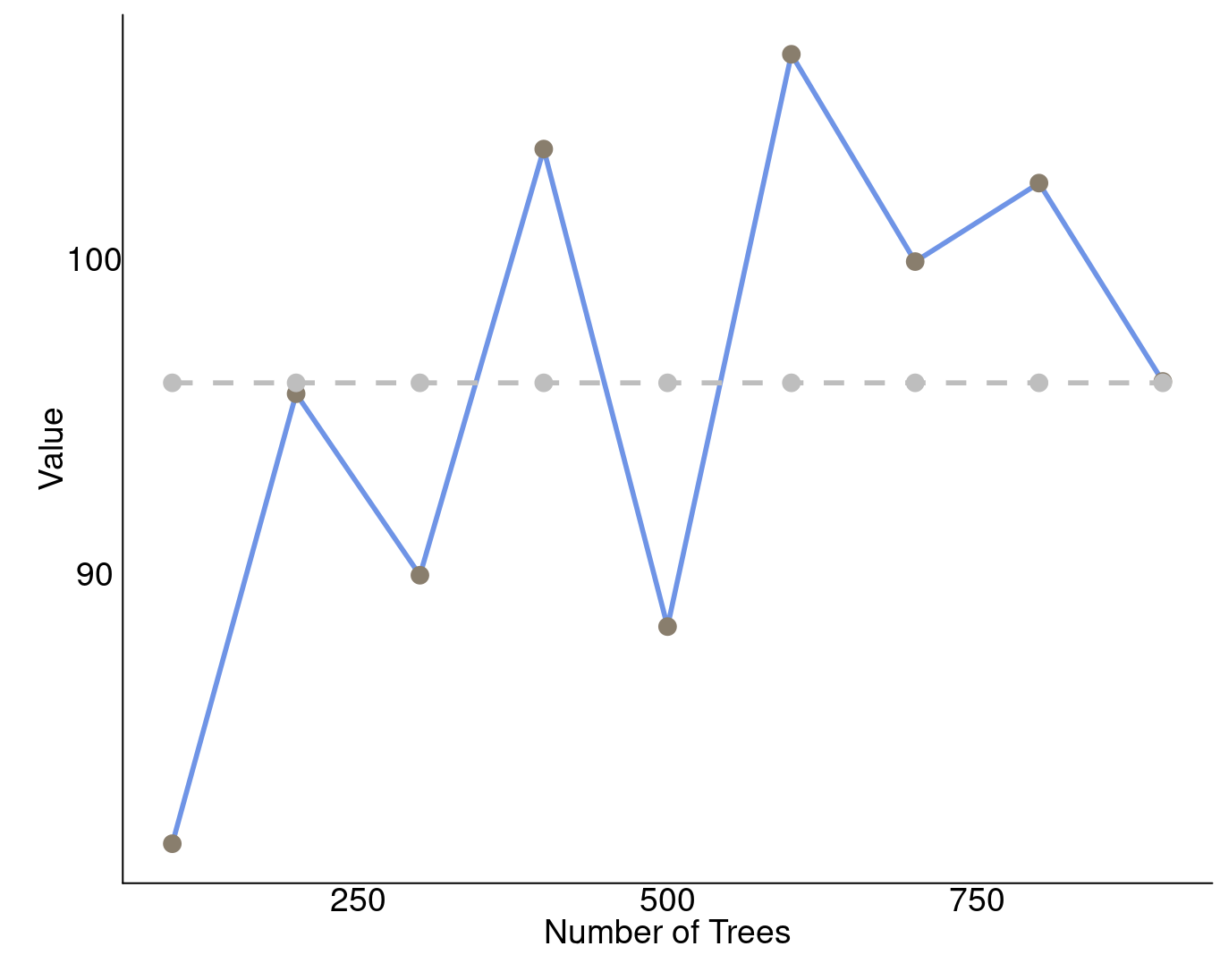}
    \caption{Performance of DTR estimator by number of trees (1 strata)}
    \label{fig:val_1_strat}
\end{figure}

We first plot the performance against the number of trees, as a form of hyperparameter tuning. In Figure \ref{fig:val_1_strat}, the blue line represents the performance of our method, and the grey line represents the value of the observed policy, which is what patients actually received. Performance of our method in the $1$ strata case peaks with $600$ trees.

We can plot the results of the analysis using $600$ trees and can see that although the estimated policy outperforms the observed policy, this only extends survival by $\sim 10$ days, and does not perform as well as the 2 strata scenario, which extended survival by $\sim 30$ days.

\begin{figure}[h]
    \centering
    \includegraphics[width=0.7\linewidth]{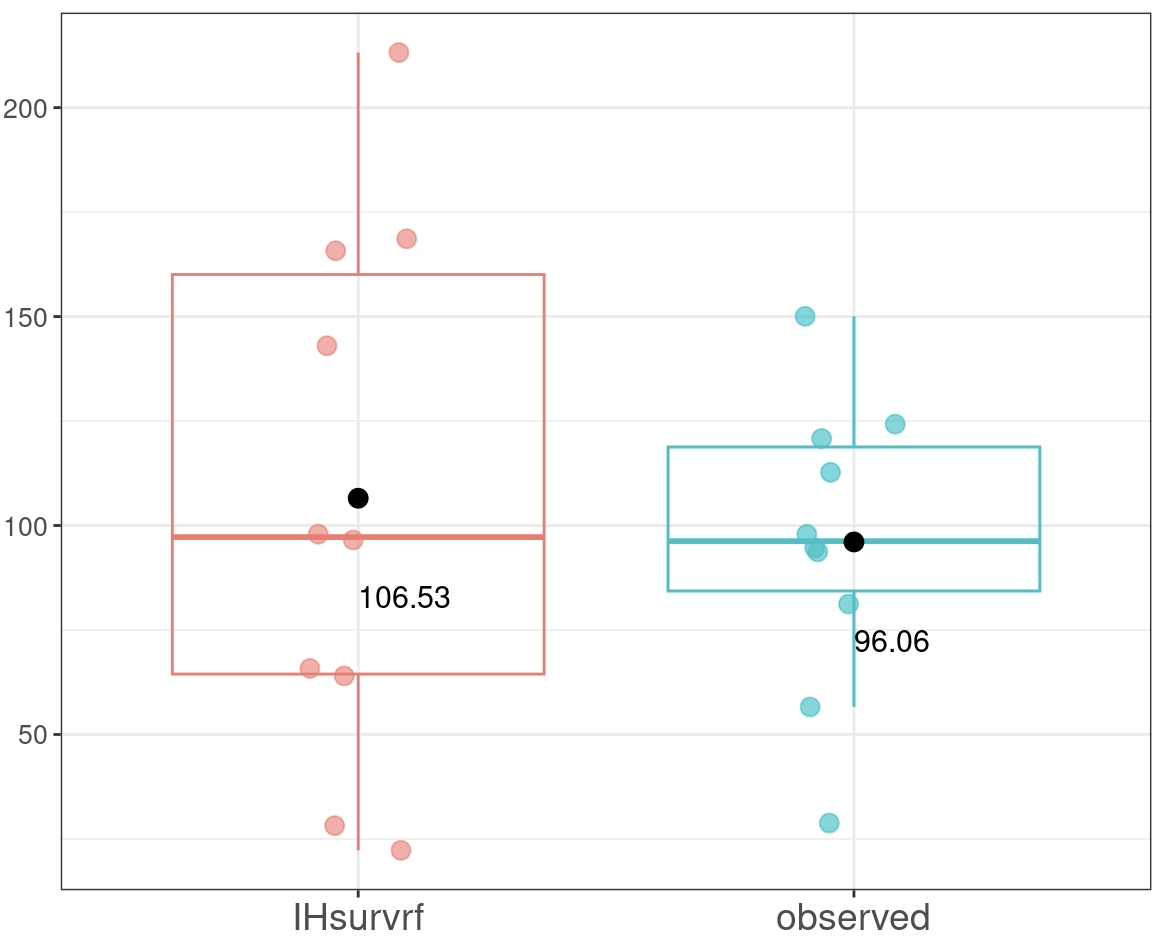}
        \caption{Performance of DTR estimator using 600 trees)}
\end{figure}

%% file: main.bbl
\begin{thebibliography}{}

\bibitem[Barnes et~al., 2021]{barnes2021incidence}
Barnes, E.~L., Herfarth, H.~H., Kappelman, M.~D., Zhang, X., Lightner, A., Long, M.~D., and Sandler, R.~S. (2021).
\newblock Incidence, risk factors, and outcomes of pouchitis and pouch-related complications in patients with ulcerative colitis.
\newblock {\em Clinical Gastroenterology and Hepatology}, 19(8):1583--1591.

\bibitem[Barnes et~al., 2020]{barnes2020decreasing}
Barnes, E.~L., Jiang, Y., Kappelman, M.~D., Long, M.~D., Sandler, R.~S., Kinlaw, A.~C., and Herfarth, H.~H. (2020).
\newblock Decreasing colectomy rate for ulcerative colitis in the united states between 2007 and 2016: a time trend analysis.
\newblock {\em Inflammatory bowel diseases}, 26(8):1225--1231.

\bibitem[Blatt et~al., 2004]{blatt2004learning}
Blatt, D., Murphy, S.~A., and Zhu, J. (2004).
\newblock A-learning for approximate planning.
\newblock {\em Ann Arbor}, 1001:48109--2122.

\bibitem[Chakraborty and Murphy, 2014]{chakraborty2014dynamic}
Chakraborty, B. and Murphy, S.~A. (2014).
\newblock Dynamic treatment regimes.
\newblock {\em Annual review of statistics and its application}, 1:447--464.

\bibitem[Cho et~al., 2023]{cho2023multi}
Cho, H., Holloway, S.~T., Couper, D.~J., and Kosorok, M.~R. (2023).
\newblock Multi-stage optimal dynamic treatment regimes for survival outcomes with dependent censoring.
\newblock {\em Biometrika}, 110(2):395--410.

\bibitem[Cho et~al., 2022]{cho2022interval}
Cho, H., Jewell, N.~P., and Kosorok, M.~R. (2022).
\newblock Interval censored recursive forests.
\newblock {\em Journal of Computational and Graphical Statistics}, 31(2):390--402.

\bibitem[Clifton and Laber, 2020]{clifton2020q}
Clifton, J. and Laber, E. (2020).
\newblock Q-learning: Theory and applications.
\newblock {\em Annual Review of Statistics and Its Application}, 7:279--301.

\bibitem[Ertefaie, 2014]{ertefaie2014constructing}
Ertefaie, A. (2014).
\newblock Constructing dynamic treatment regimes in infinite-horizon settings.
\newblock {\em arXiv preprint arXiv:1406.0764}.

\bibitem[Goldberg and Kosorok, 2012]{goldberg2012q}
Goldberg, Y. and Kosorok, M.~R. (2012).
\newblock Q-learning with censored data.
\newblock {\em Annals of statistics}, 40(1):529.

\bibitem[Hernan, 2024]{Hernan2024-HERCIW}
Hernan, M.~A. (2024).
\newblock {\em Causal Inference: What If}.
\newblock Taylor \& Francis, Boca Raton.

\bibitem[Huang et~al., 2014]{huang2014optimization}
Huang, X., Ning, J., and Wahed, A.~S. (2014).
\newblock Optimization of individualized dynamic treatment regimes for recurrent diseases.
\newblock {\em Statistics in medicine}, 33(14):2363--2378.

\bibitem[Ishwaran et~al., 2008]{ishwaran2008random}
Ishwaran, H., Kogalur, U.~B., Blackstone, E.~H., and Lauer, M.~S. (2008).
\newblock Random survival forests.

\bibitem[Jiang et~al., 2017]{jiang2017estimation}
Jiang, R., Lu, W., Song, R., and Davidian, M. (2017).
\newblock On estimation of optimal treatment regimes for maximizing t-year survival probability.
\newblock {\em Journal of the Royal Statistical Society Series B: Statistical Methodology}, 79(4):1165--1185.

\bibitem[Kosorok, 2008]{kosorok2008introduction}
Kosorok, M.~R. (2008).
\newblock {\em Introduction to empirical processes and semiparametric inference}, volume~61.
\newblock Springer.

\bibitem[Kosorok and Laber, 2019]{kosorok2019precision}
Kosorok, M.~R. and Laber, E.~B. (2019).
\newblock Precision medicine.
\newblock {\em Annual review of statistics and its application}, 6:263--286.

\bibitem[Long et~al., 2013]{long2013increased}
Long, M.~D., Martin, C., Sandler, R.~S., and Kappelman, M.~D. (2013).
\newblock Increased risk of herpes zoster among 108 604 patients with inflammatory bowel disease.
\newblock {\em Alimentary pharmacology \& therapeutics}, 37(4):420--429.

\bibitem[Luckett et~al., 2019]{luckett2019estimating}
Luckett, D.~J., Laber, E.~B., Kahkoska, A.~R., Maahs, D.~M., Mayer-Davis, E., and Kosorok, M.~R. (2019).
\newblock Estimating dynamic treatment regimes in mobile health using v-learning.
\newblock {\em Journal of the American Statistical Association}.

\bibitem[Meinshausen and Ridgeway, 2006]{meinshausen2006quantile}
Meinshausen, N. and Ridgeway, G. (2006).
\newblock Quantile regression forests.
\newblock {\em Journal of machine learning research}, 7(6).

\bibitem[Moodie et~al., 2012]{moodie2012q}
Moodie, E.~E., Chakraborty, B., and Kramer, M.~S. (2012).
\newblock Q-learning for estimating optimal dynamic treatment rules from observational data.
\newblock {\em Canadian Journal of Statistics}, 40(4):629--645.

\bibitem[Murphy, 2005]{murphy2005generalization}
Murphy, S.~A. (2005).
\newblock A generalization error for q-learning.

\bibitem[Rubin, 2005]{rubin2005causal}
Rubin, D.~B. (2005).
\newblock Causal inference using potential outcomes: Design, modeling, decisions.
\newblock {\em Journal of the American Statistical Association}, 100(469):322--331.

\bibitem[Schulte et~al., 2014]{schulte2014q}
Schulte, P.~J., Tsiatis, A.~A., Laber, E.~B., and Davidian, M. (2014).
\newblock Q-and a-learning methods for estimating optimal dynamic treatment regimes.
\newblock {\em Statistical science: a review journal of the Institute of Mathematical Statistics}, 29(4):640.

\bibitem[Simoneau et~al., 2020]{simoneau2020estimating}
Simoneau, G., Moodie, E.~E., Nijjar, J.~S., Platt, R.~W., Investigators, S. E. R. A. I.~C., et~al. (2020).
\newblock Estimating optimal dynamic treatment regimes with survival outcomes.
\newblock {\em Journal of the American Statistical Association}, 115(531):1531--1539.

\bibitem[Song et~al., 2015]{song2015penalized}
Song, R., Wang, W., Zeng, D., and Kosorok, M.~R. (2015).
\newblock Penalized q-learning for dynamic treatment regimens.
\newblock {\em Statistica Sinica}, 25(3):901.

\bibitem[Watkins and Dayan, 1992]{watkins1992q}
Watkins, C.~J. and Dayan, P. (1992).
\newblock Q-learning.
\newblock {\em Machine learning}, 8:279--292.

\bibitem[Ye et~al., 2023]{ye2023stage}
Ye, H., Zhou, W., Zhu, R., and Qu, A. (2023).
\newblock Stage-aware learning for dynamic treatments.
\newblock {\em arXiv preprint arXiv:2310.19300}.

\bibitem[Zhang et~al., 2015]{zhang2015using}
Zhang, Y., Laber, E.~B., Tsiatis, A., and Davidian, M. (2015).
\newblock Using decision lists to construct interpretable and parsimonious treatment regimes.
\newblock {\em Biometrics}, 71(4):895--904.

\bibitem[Zhao et~al., 2011]{zhao2011reinforcement}
Zhao, Y., Zeng, D., Socinski, M.~A., and Kosorok, M.~R. (2011).
\newblock Reinforcement learning strategies for clinical trials in nonsmall cell lung cancer.
\newblock {\em Biometrics}, 67(4):1422--1433.

\bibitem[Zhao et~al., 2015]{zhao2015new}
Zhao, Y.-Q., Zeng, D., Laber, E.~B., and Kosorok, M.~R. (2015).
\newblock New statistical learning methods for estimating optimal dynamic treatment regimes.
\newblock {\em Journal of the American Statistical Association}, 110(510):583--598.

\bibitem[Zhou et~al., 2024a]{zhou2024optimal}
Zhou, C.~W., Freeman, N.~L., McGinigle, K.~L., and Kosorok, M.~R. (2024a).
\newblock Optimal individualized treatment regimes for survival data with competing risks.
\newblock {\em arXiv preprint arXiv:2411.08315}.

\bibitem[Zhou et~al., 2024b]{zhou2024estimating}
Zhou, W., Zhu, R., and Qu, A. (2024b).
\newblock Estimating optimal infinite horizon dynamic treatment regimes via pt-learning.
\newblock {\em Journal of the American Statistical Association}, 119(545):625--638.

\end{thebibliography}
